
---------------------------------------------------------------

\magnification 1200
\baselineskip 16pt
\def\Oc{{\cal O}}
\def\Dc{i{\cal D}}
\def\Dcs{i\!\thru{{\cal D}}}
\def\Dct{i\tilde{\cal D}}
\def\Dcst{i\!\thru{\tilde{\cal D}}}
\def\D{{\bf D}}
\def\Db{\bar{\bf D}}
\def\X{{\bf X}}
\def\Y{{\bf Y}}
\def\A{{\bf A}}
\def\J{{\bf J}}
\def\P{{\bf P}}
\def\M{{\bf M}}
\def\thru#1{\mathrel{\mathop{#1\!\!\!/}}}
\def\thruv#1{\mathrel{\mathop{#1\!\!\!\!/}}}
\def\underarrow#1{\mathrel{\mathop{\longrightarrow}\limits_{#1}}}
\def\ds{i\!\thru\partial}
\def\Tr{\hbox{\rm Tr}\,}
\def\tr{\hbox{\rm tr}}
\def\Det{\hbox{\rm Det}\,}
\def\Re{\hbox{\rm Re}\,}
\def\Im{\hbox{\rm Im}\,}
\def\dd{\hbox{\rm d}}
\def\Dzero{\hbox{{\bf D}}_0}
\def\bra{\langle}
\def\ket{\rangle}
\def\Bra{\left\langle}
\def\Ket{\right\rangle}

\null
\centerline{\bf WIGNER TRANSFORMATION }
\centerline{\bf FOR THE DETERMINANT OF DIRAC OPERATORS}
\vskip1cm
\centerline{ L.L. Salcedo${}^1$ and E. Ruiz Arriola${}^{1,2}$ }
\vskip0.5cm
\centerline {${}^1$Departamento de F\'{\i}sica Moderna, Universidad de Granada}
\centerline {E-18071 Granada, Spain }
\vskip0.5cm
\centerline{ ${}^2$National Institute for Nuclear Physics and High Energy
Physics, (NIKHEF-K)}
\centerline{ 1009-DB Amsterdam, The Netherlands}
\vskip1cm
\centerline{\bf ABSTRACT}
\vskip 18pt
\vbox{
\baselineskip 12pt
\narrower{
We use the $\zeta$-function regularization and an integral representation
of the complex power of a pseudo differential operator, to give an
unambiguous definition of the determinant of the Dirac operator.
We bring this definition to a workable form by making use of an
asymmetric Wigner representation. The expression so obtained is amenable
to several treatments of which we consider in detail two, the inverse mass
expansion and the gradient expansion, with concrete examples.
We obtain explicit closed expressions for the corresponding Seeley-DeWitt
coefficients to all orders. The determinant is shown to be vector gauge
invariant and to posses the correct axial and scale anomalies.
The main virtue of our approach is that
it is conceptually simple and systematic and can be extended naturally
to more general problems (bosonic operators, gravitational fields, etc).
In particular, it avoids
defining the real and imaginary parts of the effective action
separately. In addition, it does not reduce the problem to a bosonic one to
apply heat kernel nor
performs further analytical rotations of the fields to make the Dirac
operator Hermitian.
We illustrate the flexibility of the method by studying some interesting cases.

}}
\vskip1cm
\centerline {\sl {15 December 1994}}
\vfill
\hbox{UG-DFM-33/94 \hfill}
\eject
\null\medskip
{\bf 1. Introduction}
\medskip
The success of the Lagrangian formulation of relativistic field theories
is due to the easy implementation of symmetries and in particular of Lorentz
invariance. The same is true for the associated covariant
quantization procedure,
the Feynman functional integral approach. In a typical setting, as will
be ours in this work, one considers Dirac fermions in $D$ space-time
dimensions in the presence of arbitrary non gravitational bosonic
background fields coupled to Lorentz and to internal symmetry indices
of the fermions. Generally speaking,
the background fields can be treated as external.
This is no restriction since they can be quantized introducing the
corresponding functional integration over them.
The effective action of the system is then obtained by integrating out
the fermion fields. Formally the Grassmann integral gives the determinant
of the Dirac operator, $\Det\D$, and the effective action is just
its logarithm. As a consequence this determinant
plays an important role in the functional integral formulation [1].

In the relativistic case the naive determinant is ultraviolet divergent
and one has to define a renormalized determinant by introducing counterterms
or some other equivalent technique. In this process some classical symmetries
can be lost. Of course, all this parallels the diagrammatic approach where
the determinant is represented by the one fermion loop graphs.

There has been a number of ways to address the problem of defining a finite
determinant or equivalently a finite effective action [2].
Most of the
methods developed in the literature try to reduce the problem to a
bosonic one. This is because second order differential
operators are algebraically
simpler and better studied in this context. In addition, the important
ultraviolet problem can be treated within an inverse mass expansion
(i.e. a simultaneous weak and smooth field expansion) with the heat kernel
technique, and this applies to second order definite positive Hermitian
differential
operators. The reduction is usually achieved by considering $\D^2$ or
$\D\D^\dagger$ and $\D^\dagger\D$ [3-8].
In the first case a further analytical
rotation of the fields (besides the Euclidean rotation) is assumed to make
$\D$ antihermitian. This  requires to extend the internal symmetry group too
and it is known that the rotation back of concrete subgroups can be ambiguous
[2].
Furthermore, although $\D^2$ looks like a bosonic theory, some of the
efficient methods developed to go beyond the inverse mass expansion [9],
may not be straightforwardly applied to fermions.
This is because such methods assume
the unrestricted validity of the formal relation $\Det(AB)=\Det(A)\Det(B)$,
which does not hold under regularization. In the bosonic case the use of
this formal relation just redefines the counterterms, but for fermions it makes
the determination of the chiral anomaly ambiguous.
Other approaches use formal relations
to define independently the real and imaginary parts of the effective action,
using $\D$ and $\D^\dagger$ and are rather ad hoc [8,10].
Similar approaches used to define an ad hoc Jacobian of the functional measure
under chiral transformations are known to lead to results
inconsistent with the Bardeen anomaly [11].

A different approach is that of Leutwyler [12] and Ball [2].
It uses a formal definition of the
variation of the effective action in Euclidean space
which is regularized in a chirally
invariant way using $\D^\dagger\D$ in a proper time representation.
For the real part of the action this defines a true variation.
The regularized variation of
the imaginary part satisfies the integrability condition only after
adding suitable polynomial counterterms which introduce the chiral anomaly.
This approach is both mathematically impeachable and also computationally
convenient in heat kernel-like expansions. However it is rather sophisticated
and relies heavily on the previous knowledge of the subtleties involved
in $\Det\D$ obtained after years of deeply original insights [13-20].
For instance the operator
$\D^\dagger$, which in principle is unrelated to the problem and formally
cancels in the definition, is needed in the construction.
It is not clear to us how this approach can be properly extended to more
general theories, such as non local theories, string theories, and so on.

We think that it can be of interest to pursue an alternative statement
of the problem where the definition of the determinant is given at the
very beginning and then every other quantity can be defined and calculated
unambiguously without the need of new prescriptions.
A good example of the latter is provided by the functional Jacobian under
variations of $\D$. Although such an object is not needed in this kind
of approach,
it is perfectly well defined and can be computed if desired [21].
The only remaining freedom is of course the addition of counterterms,
polynomials of degree $D$ in the number of derivatives plus external fields,
which allows to reproduce any other renormalization prescription
or enforce particular symmetries. A suitable definition, given long ago
[22], is the $\zeta$-function regularization which has the advantage
of preserving automatically a large class of the classical symmetries,
namely those which are implemented by similarity transformations of $\D$.
This includes relativistic invariance, vector gauge invariance
and so on. Other virtue
is that it is well grounded mathematically [23]. Using a typical
Cauchy integral representation, the $\zeta$-function can be related to
the resolvent of $\D$ which is a more tractable object and is suitable
to perform systematic expansions. Finally, a Wigner representation
[24,25] allows
us to treat properly the ultraviolet divergences appearing in the diagonal
elements of the resolvent. In this way a well defined workable form for
the effective action and other quantities is obtained. Although
by no means is it intended in this work to achieve strict mathematical rigor,
the troublesome ultraviolet sector is treated carefully, and only in the
infrared we proceed formally by assuming that $\D$ is well behaved in this
sector, that is, effective boundaryless boundary conditions and no zero
modes.

Because the subject has been extensively studied in the past, we do not
intend to present truly new theorems, rather our emphasis is on introducing
a conceptually simple scheme to make the subject more easily graspable with
more systematic, and sometimes simpler, proofs of known results and
also with an eye put on the generalization to worse known systems where
our intuition is less developed.

The paper is organized as follows. In Section 2, we present our conventions
and set up the basic formalism for the $\zeta$-function regularized determinant
of the Dirac operator, as well as the corresponding consistent currents.
Actually, we find that the effective action can be reconstructed from the
current without loss of information in this particular regularization.
Section 3 deals with one special definition of the Wigner transformation,
particularly adapted to the Dirac operator, and rather convenient
from a computational point of view.
In Section 4 we consider an inverse mass expansion for the Dirac operator
and among other things, we obtain a direct, i.e. non recursive, determination
of the corresponding Seeley-DeWitt coefficients. These can be profitably
used to write down inverse mass expansions both for the effective action
and the effective currents. A further application of the previous results
can be found in Section 5, when
computing chiral and scale anomalies within the $\zeta$-function
regularization. Moreover, we establish the general form of the counterterms
needed to bring the chiral anomaly to its minimal (Bardeen) form. To do
so the Wigner transformation method turns out to be very useful.
In Section 6, some aspects of the so-called gradient or derivative expansion
are studied within the present formalism, i.e. Wigner transformation technique.
We consider for instance a direct calculation of the gauged
Wess-Zumino-Witten term from the Dirac determinant.
We do so without using better known methods, like e.g. trial and error,
differential geometry or integration of the chiral anomaly.
Along similar lines, we study fermionic currents and their relation
to effective actions in two less dimensions, again with the help of the Wigner
transformation method. Finally, parity anomalies in odd dimensions are
revisited in Section 7, where the gauge invariant Chern-Simons action
is obtained. In Appendix A, we establish a explicit closed formula for the
Seeley-DeWitt coefficients of the Dirac operator to all even orders.
In Appendix B, the connection between the well-known Heat Kernel expansion
for the squared Dirac operator, $\D^2$,
and our inverse mass expansion for the Dirac operator, $\D$,
is developed in the even-dimensional case.
Appendices C and D collect explicit formulas for the chiral anomaly and
the counterterms for the most general Dirac operator in four dimensions.

\medskip
{\bf 2. $\zeta$-function regularization}
\medskip

Let $\D$ be the Dirac operator in $D$ dimensional Euclidean space for a
fermion in the presence of arbitrary (non gravitational) external fields
$$ \D = \ds +\Y(x) \eqno(2.1)$$
Here $\Y(x)$ is a matrix in the internal degrees of freedom, i.e.
spinor and flavor, but does not contain derivative operators.
Our conventions regarding gamma matrices are as follows
$$\eqalign{
\gamma_\mu^\dagger &= -\gamma_\mu\,,\qquad
\{ \gamma_\mu , \gamma_\nu \} = - 2\delta_{\mu\nu} \cr
\gamma_5 &= -i^{D/2}\gamma_0\gamma_1\cdots\gamma_{D-1}\,,\qquad
\epsilon_{01\dots D-1}=+1 \cr
}\eqno(2.2)$$

Whenever needed we will assume the standard hermiticity for the external
fields,
that is, such that if they transform covariantly under the Wick rotation,
$\gamma_0\D$ is Hermitian in Minkowski space. This implies that the
Euclidean effective action is real in the pseudoparity even sector
(containing no Levi-Civita pseudotensor)
and imaginary in the pseudoparity odd one (containing a
 Levi-Civita pseudotensor).
We will often use the object $\D$ itself rather than $\Y(x)$
because this produces more compact formulas. Another important
point is that $\D$ transforms homogeneously under the classical symmetry
transformations
$$\eqalign{\D\to\D^\Omega &= \Omega_2\D\Omega_1^{-1} \cr
 \psi^\Omega(x) &= \Omega_1\psi(x^\prime) \cr
 \bar\psi^\Omega(x) &= \bar\psi(x^\prime)\Omega_2^{-1} \cr }
\eqno(2.3)$$
whereas $\Y$, in general, transforms inhomogeneously.
This transformation corresponds to
a classical symmetry if it maintains the structure (2.1), i.e. $\D^\Omega=
\ds + \Y^\Omega(x)$, but otherwise $\Omega_1$, and $\Omega_2$ can depend on
$x$ and contain derivative operators. In this paper we will consider
explicitly two classical symmetries, namely
chiral gauge rotations and scale transformations. Notice that the form
of the Dirac operator does not include general coordinate
transformations as classical symmetries. This would require
to extend the Dirac
operator by properly coupling gravitational fields, and subsequent
generalization of our computational procedure to curved space-time.
Such a study will not be undertaken here and is left for future research.

For definiteness we can think of $\D$ as admitting a complete set of
left and right eigenvectors
$$\D\phi_n(x) = \lambda_n\phi_n(x)\,,\qquad\xi_n(x)\D =
\lambda_n\xi_n(x) \eqno(2.4)$$
which can be normalized (in a box) so that $\langle\xi_n|\phi_k\rangle
=\delta_{nk}$. More generally $\D$ can have a general Jordan form, i.e.
completeness of the eigenvectors will not be required.
The Euclidean partition function is a functional of the external
fields $\Y(x)$ given by
$$Z = \int {\cal D}\bar\psi{\cal D}\psi \exp \left\{
-\int \dd^Dx\,\bar\psi(x)\D\psi(x)\right\} \eqno(2.5)$$
and formally $Z = \Det\D$. Hence the Euclidean effective action
is formally
$$W=-\log\,Z = -\Tr\log\,\D = -\sum_n\log(\lambda_n)  \eqno(2.6)$$
The sum in (2.6) is ultraviolet divergent and must be
regularized. To do so we shall adopt the $\zeta$-function regularization
prescription [1,22]. We shall assume that $\D$ has no eigenvalues
in some neighborhood of zero,
otherwise some infrared regularization would be needed too. If there is
only a finite number of zero modes, isolated from the rest of the spectrum,
one can define a restricted determinant excluding the zero modes.

Let us consider the pseudo differential operator $(\D/\mu)^s$, where
$\mu$ is a scale introduced for dimensional bookkeeping and the
complex number $s$ is the regulator. If $\D$ admits a complete
set of eigenvectors, this operator is characterized by its eigenvectors
and eigenvalues $\{\phi_n,(\lambda_n/\mu)^s\}$. In any case a
convenient representation is given by [23]
$$ \Big({\D\over\mu}\Big)^s = -\int_\Gamma{\dd z\over 2\pi i}
\Big({z\over\mu}\Big)^s {1\over \D-z} \eqno(2.7)$$
where $\Gamma$ is a path that starts at infinity, follows a ray of
minimal growth (i.e. no eigenvalue of $\D$ lies on it), encircles the
origin clockwise and goes back to infinity along the ray. No
eigenvalues are encircled by $\Gamma$. If the set of eigenvalues is
bounded, $\Gamma$ can be deformed to an anticlockwise oriented closed path
containing the eigenvalues of $\D$ but excluding the origin.
This construction assumes that no eigenvalue lies in a neighborhood of
the origin and hence that $\D$ is non singular.
This representation is meaningful for $\Re s < 0$, and can be
analytically continued to the complex plane of $s$. With this
prescription, $\bra x|\D^s|y\ket$ is an entire function of $s$ for
$x\ne y$ and has simple poles at $s=-1,-2,\dots,-D$, for $x=y$ [23].
(Here $|x\ket$ is an eigenvector of the multiplicative operator
$\hat{x}_\mu$ in $D$-dimensional Euclidean space. In what follows
we will not use a different notation for a function of $x_\mu$ and the
corresponding multiplicative operator, and similarly we will identify
the momentum operator and $i\partial_\mu$). In particular $\D^s$ is
analytic at $s=0$ and this allows for the definition of a regularized
logarithm of $\D$
$$\log\Big({\D\over\mu}\Big) = {\dd\over\dd
s}\Big({\D\over\mu}\Big)^s\Big|_{s=0} \eqno(2.8)$$
and of the regularized determinant of $\D$ [1,22]
$$W(\D) = -\log\Det\Big({\D\over\mu}\Big)=
-\Tr\log\Big({\D\over\mu}\Big) = -{\dd\over\dd s}
\int\dd^Dx\,\tr\bra
x|\Big({\D\over\mu}\Big)^s|x\ket\Big|_{s=0}\eqno(2.9)$$
The determinant so defined is completely finite but depends on the
arbitrary scale $\mu$ introduced by the regularization.

Using previous formulas it is easy to prove the following identity
satisfied by the action in this regularization
$$W(\D) = -{\dd \over \dd s}{\Gamma(s+1)\over \Gamma(s+n)}
\int_\Gamma{\dd z\over 2\pi i} z^{s+n-1}(-{\partial\over\partial z})^n
W(\D-z)|_{s=0}\eqno(2.10) $$
In fact in the r.h.s. we can use any other regularization because all of them
differ by a polynomial in $z$ and its contribution cancels in the integral.
On the other hand using a value $n>D$, $\partial^n_z W(\D-z)$ is ultraviolet
finite and independent of the regularization and can be used to reconstruct
the action.

It is also of interest to obtain expressions for the variation of
the action under a generic infinitesimal transformation of the
fields in the Dirac operator. Let $\delta_X\D = \X$ be such a
variation of $\D$, where the only restriction is that $\X$ is a
multiplicative operator, that is, without derivatives. Hence it corresponds
to an infinitesimal classical symmetry transformation. Correspondingly
$$\delta_X\D = \X, \quad\quad \delta_X W = \int\dd^Dx\,\tr(\X\J),\quad\quad
\J={\delta W\over\delta\D} \eqno(2.11)$$
We will refer to $\J$ as the current associated to $W$, although
usually this name is reserved for the variation under
gauge fields and otherwise $\J$ is called a density. The current
so defined satisfies consistency conditions [17] which reflect that it is
a true variation and hence it is known as the consistent current.
Other definitions of the current are sometimes more convenient,
in particular the so called (chiral) covariant current [26],
which will be considered in Section 5. They differ by a polynomial in the
fields and derivatives.

An explicit expression for the current can be obtained
by using the technique introduced in [21,27]
$$\delta_X W = -\delta_X{\dd\over\dd s}\Tr {\D^s\over\mu^s}\Big|_{s=0}
= -{\dd\over\dd s}s\Tr(\X{\D^{s-1}\over\mu^s})\Big|_{s=0} \eqno(2.12)$$
We have used the cyclicity of the trace in presence of the
regulator $s$. Because $\X$ is local, we have
$$\J(x;\D) = -{\dd\over\dd s} s\bra x|{\D^{s-1}\over\mu^s}
|x\ket\Big|_{s=0} \eqno(2.13)$$
The current can be used to recover the action by applying Eq. (2.10) for
$n=1$,
$$W(\D) = -\int\dd^Dx\,\tr{\dd \over \dd s}
\int_\Gamma{\dd z\over 2\pi i} z^s \J(x;\D-z)|_{s=0}\eqno(2.14) $$
Again, we can use any other regularization for the current in the r.h.s.,
moreover there is no need to use the consistent current, it can be for
instance the covariant current, and the $z$ integral will
take care of introducing the proper chiral anomaly into the action.

The $\zeta$-function regularized action enjoys all the classical
symmetries which are also symmetries of the quantum Dirac equation i.e. the
eigenvalue equation $\D\phi_n=\lambda_n\phi_n$. This includes in particular
vector gauge invariance, $\D\to\Omega(x)\D\Omega^{-1}(x)$, with
$\Omega(x)$ a matrix valued function acting on flavor space but not in
Dirac space. On the other hand scale and axial transformations
are symmetries only of the classical equation $\D\phi=0$, and the
corresponding currents are anomalous. Note that the classical equation
does not imply a zero mode in the quantum equation due to the
different boundary conditions of both equations. Actually the solutions
of the classical and the quantum equations
are normalizable in a spatial box and space-time box respectively.
The scale and axial anomalies will be considered in Section 5.

\medskip

{\bf 3. Wigner transformation}
\medskip
Seeley's representation (2.7)
requires to invert the operator $\D-z$.  This can be conveniently
accomplished by means of an asymmetric version of the Wigner
representation [24,25].  For any operator $A$, let
$$A(x,p) =
\int\dd^Dy\,e^{iyp}\bra x|A|x-y\ket = {\bra x|A|p\ket\over
\bra x|p\ket} \eqno(3.1)$$
be its (asymmetric) Wigner representation.  $|p\ket$ is
the momentum eigenstate with $\bra x|p\ket = e^{-ixp}$.
When $A$ is a pseudo differential operator [28], the
quantity so defined is closely related to the so called symbol
of $A$. For a pseudo differential operator $A$ of complex order $m$,
$A(x,p)$ is of the form $\sum_{j=0}^\infty a_{m-j}(x,p)$ where
$a_k(x,p)$ are homogeneous functions of degree $k$ in $p$. In
particular $\D^s$ is of order $s$ [23].

{}From this definition
$$\bra x|A|x\ket = \int{\dd^Dp\over(2\pi)^D}
A(x,p)\,,\quad \Tr A = \int{\dd^Dx\dd^Dp\over(2\pi)^D} \tr A(x,p)\,,
\eqno(3.2)$$
where $\tr$ acts on internal and Dirac spinor degrees of freedom only, and the
product of two operators satisfy the following formula
$$(AB)(x,p) =
\exp(i\partial_p^A\cdot\partial_x^B)A(x,p)B(x,p)\,,  \eqno(3.3)$$
where $\partial^A_p$ acts only on the $p$-dependence in $A(x,p)$ and
$\partial_x^B$ on the $x$-dependence in $B(x,p)$.

Let the propagator or resolvent of $\D$, be
$$G(z)=(\D -z)^{-1} \eqno(3.4)$$
and $G(x,p;z)$ its Wigner representation.
A convenient expression for $G(x,p;z)$ can be obtained using the
trick of ref. [9]. Recalling $\hat{x}_\mu|x\rangle = x_\mu|x\rangle$
$$\eqalign{
\langle x|G(z)|p\rangle &= \langle x|e^{-ik\hat{x}}e^{+ik\hat{x}}
(\D -z)^{-1}|p\rangle \cr
 &= e^{-ikx}\langle x|(\thru{k}+\D -z)^{-1}e^{+ik\hat{x}}|p\rangle
 = \langle x|k\rangle \langle x|(\thru{k}+\D -z)^{-1}|p-k\rangle \cr
}\eqno(3.5) $$
where we have made use of $e^{+ik\hat{x}}i\partial_\mu e^{-ik\hat{x}}
=i\partial_\mu +k_\mu$. Choosing $k=p$ we obtain the following
compact expression for the Wigner representation of the propagator
$$G(x,p;z) = \langle x|(\thru p+\D-z)^{-1} |0\rangle \eqno(3.6)$$
where $|0\rangle$ is the state of zero momentum, $\langle x|0\rangle = 1$.
In practice this implies that $i\partial_\mu$ derivates every
$x$ dependence at is right, until it annihilates $|0\rangle$.
At this point we have separated from the standard approach,
which uses the product formula (3.3) to set up a recurrence relation
to compute the symbol of the resolvent [21].
Our method is more efficient for it computes directly, that is non
recursively, each of the terms.

Several expansions can be devised to compute $G(x,p;z)$. Two of
them will be considered in the next sections. Let us point out
that the definition given for $A(x,p)$ is not gauge covariant because
$|p\ket$ is not. It would be very interesting to have a gauge invariant
version of the Wigner representation. In fact the problem of
finding non local covariant expansions, as required for instance
in massless theories, is still open [2, p. 128]. In what
follows we will consider only local objects of the form $\bra
x| f(\D)|x\ket $ as given by the formula
$$ \bra x| f(\D)|x\ket = -\int {\dd^Dp\over (2\pi)^D} \int_\Gamma
{\dd z\over 2\pi i} f(z) G(x,p;z) \eqno(3.7) $$
We will assume that the function $f(z)$ is sufficiently convergent at
infinity or else that it can be obtained as a suitable analytical extrapolation
from a parametric family $f(z,s)$ in the variable $s$.
In either case the integration over $z$ should be performed in the
first place, to yield the Wigner representation of the operator $f(\D)$.
Afterwards, the $p$ integration is carried out, corresponding to
take the diagonal matrix elements of $\bra y|f(\D)|x\ket$, hence restoring
gauge covariance.

Because the subject of this paper has been considered extensively in
the past [2], and to some extent it has been taken up by mathematicians,
we must make some comment on the validity of equations such as (3.2)
from a more rigorous point of view. This equation is based on an assumption,
namely that there exists an operator $\hat{x}_\mu$
such that it satisfies Heisenberg commutation relations
$[i\partial_\mu,\hat{x}_\nu]=i\delta_{\mu\nu}$. Of course this is not the
case for manifolds without a global chart and in particular for the nice
compact manifolds without boundary, usually considered in the literature.
For instance if the previous formalism is
applied to compute $\Tr(\D^s)$ with $\D = i\partial_x + m$ on the interval
$[0,T]$ compactified to a circle, we will find that the sum on the
eigenvalues of the exact formula has been replaced by an integral on a
continuum momentum label, thus introducing an approximation which
becomes exact as $T\to\infty$. On the other hand
there is no problem for R$^D$ if the compactification
comes from the nature of the external fields in the infinity. For instance
an equation similar to (3.2) for the one dimensional second order
differential operator
$A=-\partial_x^2 + x^2$ would still be exact, including detailed information
on the discrete spectrum of $A$. However such a formula would not be very
useful unless one makes some expansion, typically a gradient
expansion (to be considered in Section 6), and this kind of expansion
substitutes the discrete spectrum by a smoothed continuous density of states.
This is a common feature of any asymptotic expansion such as
heat kernel [29] or the Wigner-Kirkwood [30] expansions.

\medskip
{\bf 4. Inverse mass expansion for Dirac operators}
\medskip

{\sl 4.1 Seeley-DeWitt coefficients for Dirac operators}
\smallskip

In this section we will consider a Dirac operator $\D =\Dzero+m$,
with $m$ a mass term and $\Dzero$ of the general form (2.1). This
will allow us to obtain an expansion for the effective action in
powers of $\Dzero$ or equivalently in inverse powers of $m$. This
expansion is closely related in spirit to the Seeley-DeWitt
expansion for the heat kernel of positive definite second order
operators. Both are asymptotic expansions in the number
of external fields and their derivatives. We will consider the
quantity $\bra x|(\D/\mu)^s|x\ket$ from which the
effective action is easily obtained, Eq. (2.9). From
Eq. (3.6) it is straightforward to derive the following expansion
$$ G(x,p;z) = -\bra x|\sum_{N=0}^\infty {((\thru{p}+z-m)\Dzero)^N
(\thru{p}+z-m)\over (p^2+(z-m)^2)^{N+1}}|0\ket \eqno(4.1)$$
where recall that $\partial_\mu$ inside $\Dzero$ annihilates $|0\ket$.

The series will generally be asymptotic, giving only the analytical part of
$\bra x|(\thru{p}+m-z+\lambda\Dzero)^{-1}|0\ket$
as a function of $\lambda$ about $\lambda=0$. This expansion is
thus relevant for large $p_\mu$, $z$ and $m$ or equivalently for
sufficiently weak and smooth external fields.

Inserting the series for $G(x,p;z)$ in Eq. (3.7) for
$f(z)=(z/\mu)^s$, the following expansion will be found
$$\bra x|\left({\D\over\mu}\right)^s|x\ket = \sum_{N=0}^\infty
{1\over (4\pi)^{D/2}} {m^{s+D-N}\over\mu^s}\,c_N(s)\,\bra x|\Oc_N(\Dzero)
|0\ket
\eqno(4.2) $$
where $D$ is the space-time dimension, $c_N(s)$ is a numerical
function of $s$, $N$ and $D$, and $\Oc_N(\Dzero)$ are polynomials of
degree $N$ in $\Dzero$ which can be written in a $D$ independent
form and are matrices in internal and Dirac spinor space.
The precise factorization
between $c_N(s)$ and $\Oc_N$ is, to some extent, a matter of convenience.

To obtain explicit expressions for $\Oc_N$ it is best to work
out the numerator in Eq. (4.1) using the object
$\A_\mu ={1\over 2} \{\gamma_\mu,\Dzero\}$,
and the property $[p_\mu\A_\mu,\thru{p}]=0$.
Afterwards, an angular average over $p_\mu$ is done so that the
following integral applies:
$$\eqalign{
I_1( & k,a,b,D,s,m) = \int{\dd^Dp\over(2\pi)^D} \int_\Gamma{\dd
z\over 2\pi i} z^s p^{2k} (z-m)^a (p^2+(z-m)^2)^{-b} \cr
}\eqno(4.3)$$
where $m>0$, $\Gamma$ goes along the real negative axis, $D,k,a,b\in$Z,
$D,a \ge 0$, $D+2k>0$ and $\Re(s) < 2b - D-2k-a-1$. Explicitly
$$I_1 = {(-1)^D\over(4\pi)^{D/2}}
{\Gamma({1\over 2}D+k)\Gamma(b-{1\over 2}D-k)\Gamma(s+1) m^{s+1+D+2k+a-2b}
\over
\Gamma({1\over 2}D)\Gamma(b)\Gamma(2b-D-2k-a)\Gamma(s+2+D+2k+a-2b)}
\eqno(4.4)$$
If $m$ is allowed to be negative an additional factor $\epsilon(m)^D$ has
to be included, $\epsilon(x)$ being the sign of $x$. This circumstance
will be relevant in the odd dimensional case (see Section 7).

In this way one obtains for $c_N(s)$
$$\eqalign{
c_N(s) &= \zeta_{N-D}{\Gamma(s+1)\over\Gamma(s+1-N+D)} \cr
\zeta_{N-D} &= (-1)^D {\Gamma(\bar{N}-{1\over 2}D)\over\Gamma(N-D+1)}\,,
\qquad\qquad
\bar{N}=\lfloor{N\over 2} +1\rfloor \cr} \eqno(4.5)$$
where $\lfloor x\rfloor$ stands for largest integer not exceeding $x$.
This explicit expression for $\zeta_K$ holds for $N \ge D$ and an
analytical continuation in $D$ is understood for $N<D$. Note that
$\zeta_K$ depends also on $D$ (mod 2). These quantities can be
obtained more conveniently from the recurrence
$$\eqalign{
&  \zeta_{K-1} = \epsilon_K \zeta_K\,,\cr
& \zeta_0 = 1 \quad(\hbox{\rm even } D), \qquad
  \zeta_0 = -\Gamma({1\over 2}) \quad(\hbox{\rm odd } D),\cr
& \epsilon_K=2 \quad(\hbox{\rm even } D+K), \qquad
\epsilon_K = K \quad(\hbox{\rm odd } D+K)
\cr } \eqno(4.6)$$
In particular $\zeta_K$ vanishes for negative $K$ if $D$ is odd.
Some useful particular values of the coefficients are given in Table 4.1.

On the other hand, for the operators $\Oc_N$ at lower orders,
one obtains
$$\eqalign{
\Oc_0(\D) &= 1 \cr
\Oc_1(\D) &=  \D + \A_\mu\gamma_\mu \cr
\Oc_2(\D) &=  \D^2+\A_\mu^2 \cr
\Oc_3(\D) &=  \D^3+{1\over 2}\D^2\A_\mu\gamma_\mu
              +{1\over 2}\A_\mu\gamma_\mu\D^2+\A_\mu\D\A_\mu \cr
	&+ {1\over 3}(\A_\mu^2\A_\nu\gamma_\nu+\A_\mu\A_\nu\A_\mu\gamma_\nu
	+\A_\mu\A_\nu^2\gamma_\mu) \cr
\Oc_4(\D) &=  {1\over 2}\D^4 +
               {1\over 3}(\D^2\A_\mu^2+\A_\mu\D^2\A_\mu + \A_\mu^2\D^2)  \cr
             &+ {1\over 6} (\A_\mu^2\A_\nu^2 + (\A_\mu\A_\nu)^2 +
               \A_\mu\A_\nu^2\A_\mu) \cr }\eqno(4.7)$$
where $\A_\mu = {1\over 2}\{\gamma_\mu,\D\}$. We want to emphasize that these
expressions do not make any assumption on $\D$ other than Eqs. (2.1) and
(2.2). Note also that they have been written in a $D$ independent fashion.
The expression for the coefficients $\Oc_N(\D)$ for all even orders,
is given in Appendix A.

In the expansion (4.2), what actually appears are the matrix valued
functions $\Oc_N(x)=\bra x|\Oc_N|0\ket$, rather than the operators
$\Oc_N$ themselves.
Because the regularization is vector gauge invariant, we expect
that $\Oc_N(x)$
will be covariant under vector gauge transformations of $\D$, that is,
under $\D\to \Omega(x)\D\Omega^{-1}(x)$, $\Oc_N(x)$ will transform as
$\Oc_N(x)\to\Omega(x)\Oc_N(x)\Omega^{-1}(x)$, even if $|0\ket$
itself is not gauge invariant. Let us show this explicitly for $\Oc_2$.
Letting $\D=\Dcs + \X$ with $\Dc_\mu=i\partial_\mu + V_\mu(x)$, the
operator $\Oc_2$ can be written as
$$\Oc_2 = \X^2+{1\over 4}\{\gamma_\mu,\X\}^2 +
{1\over 2}\sigma_{\mu\nu}iF_{\mu\nu}+
{1\over 2}[\gamma_\mu,[\Dc_\mu,\X]]\,,
\quad\sigma_{\mu\nu} = {1\over 2} [\gamma_\mu ,\gamma_\nu],
\quad iF_{\mu\nu}=[\Dc_\mu,\Dc_\nu]
\eqno(4.8)$$
Therefore all the derivatives appear inside commutators and $\Oc_2$ is a
purely multiplicative operator. In this case taking the matrix element
$\bra x|\Oc_2 |0\ket$ does not break gauge invariance.
The same thing can be shown for higher orders.
A more economical manner of establishing the gauge covariance of an
expression is the following. Let $\hat{f}$ be an operator
formed algebraically out of $\Dc_\mu$ and other gauge covariant
non derivative operators $\X_i(x)$. Then of course $\hat{f}$ is itself
a gauge covariant operator.
 The matrix valued function $f(x)=\bra x|\hat{f}|0\ket$
will be gauge covariant if $\hat{f}$ is a multiplicative operator,
that is, if all the $\Dc_\mu$ in $\hat{f}$ appear as covariant
derivatives inside commutators. In turn this will be only the
case if $\hat{f}$ is invariant under $V_\mu(x)\to V_\mu(x) + a_\mu$, with
$a_\mu$ an arbitrary constant c-number, that is if
$$\delta_g\hat{f} = 0 \quad\quad \hbox{\rm under}\quad\quad
\delta_g\D = \thru{a},\quad\delta_g\X_i = 0 \eqno(4.9)$$
 For the expressions listed above one checks,
using $\delta_g\A_\mu = -a_\mu$, that in fact $\delta_g\Oc_N = 0$.
Note that this rule refers only to vector gauge invariance of local objects.
For instance in the Chern-Simons action
${1\over 16\pi}i\int\dd^3x\,\epsilon_{\mu\nu\alpha}
\,\tr(V_\mu\partial_\nu V_\alpha -{2\over 3}iV_\mu V_\nu V_\alpha)$,
(to be discussed in Section 7),
the integrand is only invariant up to a total derivative but the action itself
is invariant.

Another remark about gauge invariance is that it follows from
formal integration by parts over $p$ [9] in Eq. (3.7).
In the case of the integral $I_1(k,a,b,D,s,m)$ in Eq. (4.4), only the
relationship
$${ I_1(k+1,a-2,b,D,s,m) \over I_1(k,a,b,D,s,m) } = {{1\over
2}D+k\over b-{1\over 2}D-k-1} \eqno(4.10)$$
is required to form the gauge covariant operators $\Oc_N$.
Because this ratio is independent of $s$, the same quantities will
appear in the expansion of $\bra x|f(\D)|x\ket$ for other analytic
functions $f(z)$ whenever the $z$ integrals involved are
convergent. This argument is made rigorous in [23] for
$f(A)$, $A$ being a zeroth order pseudo differential operator
and $f(z)$ analytic in a region containing the spectrum of $A$.

It is worth noticing that the operators $\Oc_N$ are related
among themselves in a simple way. This follows from the observation
that $\D$ is invariant under
$$\delta_m\Dzero = -\delta m, \quad\quad \delta_mm=\delta m \eqno(4.11)$$
Imposing $\delta_m\bra x|\D^s|x\ket=0$ in Eq. (4.2), one finds
$$\delta_m\Oc_N = -\epsilon_{N-D}\Oc_{N-1}\delta m \eqno(4.12)$$
where the quantities $\epsilon_K$ were introduced in Eq. (4.6).
Therefore from $\Oc_N$ one can obtain algebraically
$\Oc_M,\quad M<N$. It is interesting to notice that the variations
$\delta_g$ (see Eq. (4.9)) and also $\delta_m$ (see Eq. (4.11))
greatly restrict, but do not completely fix,
the actual values of the coefficients appearing in
the expression for $\Oc_N$ as given by Eq. (4.7). It turns out
that the even order coefficients can be deduced entirely as combinatorial
factors, the odd orders being obtained by taking the variation with
respect the mass, as given by Eq. (4.12). Further details can be
looked up at Appendix A.

The coefficient $c_N(s)$ has simple poles at
$s=-1,-2,-3,\dots,N-D$ for $N<D$. This is in agreement with
general theorems for pseudo differential operators [23]. On
the other hand $c_N(s)$ vanishes for $s = 0,1,2,\dots,N-D$ if
$N \ge D$. Thus for non negative integer values of $s$, the inverse
mass expansion is exact [23] and is just a polynomial
in $m$. Setting $m=0$, one finds:
$$\bra x|\D^n|x\ket = {1\over (4\pi)^{D/2}}\zeta_n\Gamma(n+1)\Oc_{D+n}(x;\D),
\qquad\quad n=0,1,2,\dots
\eqno(4.13)$$
This result is regularization dependent, since a naive evaluation of
the matrix element would diverge.
In fact, the l.h.s. stands for $\bra x|\D^{n+s}|x\ket|_{s=0}$
through an analytical continuation in $s$,
and thus it is specific of the $\zeta$-function regularization.

{\sl 4.2 Effective action}
\smallskip

For the effective action we have, using (2.9) and (4.2),
$$\eqalign{
W(\D) &= \sum_{N=0}^\infty \left(
\alpha^W_{N-D} + \beta^W_{N-D}\log\left({m\over\mu}\right)\right)m^{D-N}
\bra\Oc_N(\Dzero)\ket \cr
&\alpha^W_K = -{\dd\over\dd s} c_{D+K}(0),\quad\quad \beta^W_K = -c_{D+K}(0)
\cr}\eqno(4.14)$$

In the r.h.s. we have used the shorthand notation
$$\bra f(\D) \ket = {1\over (4\pi)^{D/2}} \int\dd^Dx\tr\,\bra x|f(\D)|0\ket
 = {1\over (4\pi)^{D/2}} \tr\bra 0|f(\D)|0\ket \eqno(4.15)$$
The symbol $\bra\ \ \ket$ enjoys
some of the properties of the trace and in particular the trace
cyclic property, but only for multiplicative operators,
$$\bra AB\ket = \bra BA\ket \qquad\qquad\hbox{if}\quad \delta_gA
=\delta_gB=0 \eqno(4.16)$$
If one deals, however, with non multiplicative operators, like e.g.
differential operators, an additional commutator has to be added.
Also note that unlike $\Tr$, the symbol $\bra\ \ \ket$ has
dimension of $m^{-D}$.

More explicit expressions for $\alpha^W_K$ and $\beta^W_K$ are given
in Table 4.2. Note that $\alpha^W_K$, $\beta^W_K$ depend only on $D$ (mod 2).

In particular for $D=4$
$$\eqalign{
W = & \Bra
{25\over 24}m^4\Oc_0-{11 \over 9}m^3\Oc_1-{3\over 2}m^2\Oc_2
+2m\Oc_3-{1\over m}\Oc_5 + \cdots\Ket \cr
&+ \log\left({m\over\mu}\right)
\Bra-{1\over 2}m^4\Oc_0+{2\over 3}m^3\Oc_1+m^2\Oc_2-2m\Oc_3-\Oc_4
\Ket \cr }\eqno(4.17)$$

We can distinguish three contributions to the effective action,
according to their $m$ dependence, namely,
1) the contribution from $\alpha_K^W$ and $\beta_K^W\log\mu$ for $N\le D$,
which is a polynomial of degree $D$ in $m$, 2) a logarithmic
part from $\beta^W_K\log m$ with $N\le D$ and 3) an inverse mass
expansion from $\alpha^W_K$ and $N>D$. The polynomial part is
regularization dependent as it can be modified by adding
suitable local polynomial counterterms to the action. These terms
are such that the action depends on $\D$ and not on $\Dzero$ and
$m$ separately and they vanish for odd dimensions.
The logarithmic and inverse mass parts are
regularization independent. In fact by applying the operator
$(\dd/\dd m)^n$, $n>D$ at both sides of Eq.
(4.14), the l.h.s. becomes $(-1)^n\Tr(\D^{-n})$ which is
ultraviolet finite and hence independent of the renormalization
prescription. In the r.h.s. the polynomial part as well as the
dependence on $\mu$ disappears whereas the other terms remain,
yielding a pure inverse mass expansion without logarithms. The
dependence on $\log\mu$ was a trivial additive constant for the
unregularized action. This is no longer the case after
renormalization, indicating that the action has developed a
scale anomaly. The practical interest of the former arguments lies in the
possibility of reconstructing the action by dealing with explicitly convergent
and hence regularization independent objects,
namely $(-1)^n\Tr(\D^{-n})$, $n>D$, and integrating back in the mass parameter.
By properly fixing the arbitrary integration constants one might reproduce
a given renormalized action. This idea is already contained in Eq. (2.10)
for the specific case of the $\zeta$-function.

{\sl 4.3 Effective currents}\smallskip

Finally, we can also obtain an inverse mass expansion for the current
$\J$ introduced in Section 2.
Rather than computing $\J$ through the variation of $\bra\Oc_N\ket$
in Eq. (4.14), we will use the closed expression (2.13).
Recalling that $\bra x|\D^s|x\ket$ has a simple pole at $s=-1$, we
obtain the following expansion for the current
$$\J(x) = {1\over(4\pi)^{D/2}}\sum_{N=0}^\infty \left(
\alpha^J_{N-D} + \beta^J_{N-D}\log\left({m\over\mu}\right)\right)m^{D-N-1}
\Oc_N(x;\Dzero) \eqno(4.18)$$
 where
$$\eqalign{
\alpha^J_K &= -{\dd\over\dd s}(sc_{D+K}(s-1))\Big|_{s=0}
  = \epsilon_{K+1}\alpha^W_{K+1} \cr
\beta^J_K &= -sc_{D+K}(s-1)\Big|_{s=0}
  = \epsilon_{K+1}\beta^W_{K+1} \cr}\eqno(4.19)$$

Again we can check that $\delta_m\J=0$. Also comparing with the
expansion for the action, one finds
$${\delta\bra\Oc_N\ket\over\delta\D} =
{1\over(4\pi)^{D/2}} \epsilon_{N-D}\Oc_{N-1}(x)\eqno(4.20)$$
which is consistent with Eq. (4.12). A similar relation holds for the
Seeley-DeWitt coefficients [2].
It is instructive to try to obtain the same coefficients
$\Oc_N(x)$ from a heat kernel approach. The main result is that in even
dimensions, the Seeley-DeWitt coefficients of the second order bosonic
differential
operator $-\D^2$ coincide with the inverse mass expansion coefficients
of even order $\Oc_{2n}(\D)$ of the first order fermionic operator $\D$.
This result has interesting consequences and will be proven in
Appendix B.

\medskip

{\bf 5. Anomalies within the $\zeta$-function regularization}
\medskip

{\sl 5.1 Chiral and scale anomalies}\smallskip

As it is well known the effective action does not share all the
symmetries of the classical action. In the $\zeta$-function
regularization approach, this is because not all the symmetry transformations
of the classical equation $\D\phi=0$ are also symmetries of the eigenvalue
equation $\D\phi_n=\lambda_n\phi_n$. Symmetries which are broken
by a mass term classically, develop an anomaly at the quantum
level. This is the case of chiral and scale transformations.

As already mentioned under the combined set of vector and axial transformations
only the axial ones present an anomaly,
within a $\zeta$-function regularization. Vector gauge
symmetry remains unbroken. The axial anomaly is defined
[15] as the variation of the effective action
under an axial gauge transformation of $\D$, that is
$$ \D \underarrow{A} e^{-i\alpha\gamma_5}\D e^{-i\alpha\gamma_5},
\qquad \delta_A\D = -\{i\alpha\gamma_5,\D\}\,,\qquad
  {\cal A}_A = \delta_AW(\D) \eqno(5.1)$$
where $\alpha(x)$ is an infinitesimal matrix valued function
in flavor space only. This transformation is
consistent with the general structure assumed for $\D$, i.e. $\D
= \ds + \Y$, with $\Y$ a local function. ${\cal A}_A$ is the axial anomaly,
which is ultraviolet finite and of dimension
$D$ in the external fields [15].
The best way of computing ${\cal A}_A$ in the $\zeta$-function
context is to use the same trick as in Eq. (2.12):
$$\eqalign{
{\cal A}_A &= -\delta_A\Tr\log\D = -\delta_A{\dd\over\dd s}
\Tr\left(\D\over\mu\right)^s\Big|_{s=0} \cr
&={\dd\over\dd s}\Tr\left(s\{i\alpha\gamma_5,\D\}{\D^{s-1}\over\mu^s}
\right)\Big|_{s=0} = \Tr(2i\alpha\gamma_5\D^s)\Big|_{s=0} :=
\Tr(2i\alpha\gamma_5\D^0) \cr
}\eqno(5.2)$$
Here we have used the trace cyclic property and also that $\bra
x|\D^s|x\ket$ is analytic at $s=0$. Since formally $\D^0$ would correspond
to the identity operator, the previous result can be interpreted as
a regularization of this operator. Note that the anomaly is
independent of the scale $\mu$. Now using the expression for
$\bra x| \D^n|x\ket$, $n=0,1,2,\dots$ in Eq. (4.13) (for even $D$), we have
$${\cal A}_A=\bra 2i\alpha\gamma_5\Oc_D\ket \eqno(5.3)$$
which is manifestly ultraviolet finite. With the
expressions for $\Oc_2$ and $\Oc_4$ in Eq. (4.7) and after some
algebra using only defining properties of the gamma matrices, we
obtain simpler explicit forms for ${\cal A}_A$ in two and four dimensions
$$\eqalign{
(D=2)\qquad\qquad{\cal A}_A &= \bra 2i\alpha(x)\gamma_5\D^2\ket\cr
(D=4)\qquad\qquad{\cal A}_A &= \bra 2i\alpha(x)\gamma_5
({1\over 2}\D^4 + {1\over 3}\D\A_\mu^2\D)\ket\cr}\eqno(5.4)$$
These expressions are more easily arrived at
by going back to Eq. (4.1) to compute
$\Tr(2i\alpha\gamma_5\D^s)|_{s=0}$ directly. For illustration purposes
we quote in Appendix C more elaborated versions of Eq. (5.4) for a
Dirac operator with the most general spinorial structure
in two and four dimensions, after explicit evaluation of the Dirac traces.

The scale invariance can be treated in a completely similar way
in this regularization.
The trace anomaly is the variation of the effective action under
a scale transformation of $\D$ [1].
$$ \D = \ds +\Y(x)  \underarrow{S} \ds +e^{-\epsilon}\Y(e^{-\epsilon} x),
\qquad\qquad{\cal A}_S = \delta_S W \eqno(5.5)$$

We can best compute the trace anomaly by rewriting the scale transformation
as a homogeneous transformation for $\D$
$$\delta_S\D = -\epsilon (\D +[x_\mu\partial_\mu,\D]), \eqno(5.6)$$
and using the $\zeta$-function regularization of the action
$$\eqalign{
{\cal A}_S &= -\delta_S{\dd\over\dd s}
\Tr\left({\D\over\mu}\right)^s\Big|_{s=0}\cr
&=\epsilon{\dd\over\dd s}s\Tr\left((\D+[x_\mu\partial_\mu,\D])
{\D^{s-1}\over\mu^s}\right)\Big|_{s=0} =
\epsilon{\dd\over\dd s}s\Tr\left(\D{\D^{s-1}\over\mu^s}\right)\Big|_{s=0}
\cr &= \epsilon\Tr(\D^0) \cr }\eqno(5.7)$$
Likewise the axial case,
the zeroth power of $\D$ appears in the final expression for the anomaly.
Using Eq. (4.13) one finally obtains
$${\cal A}_S = \epsilon\zeta_0\bra\Oc_D\ket
 \eqno(5.8)$$
for the trace anomaly.

{}From the last line of Eq. (5.7), the scale anomaly can
alternatively be written as
$${\cal A}_S = \epsilon\mu{\partial W\over\partial\mu}
\eqno(5.9)$$
which is consistent with Eq. (4.14).
Because the axial anomaly $\delta_AW={\cal A}_A$ does not depend on $\mu$,
and using the cross variation condition $[\delta_A,\delta_S]=0$,
we conclude that ${\cal A}_S$ is chirally invariant. In addition it is also
scale invariant. Moreover the scale anomaly vanishes for odd $D$ because
from Eq. (4.20) $\delta_X\bra\Oc_D\ket = 0$ for an arbitrary variation.

Another issue is that of the anomalous breaking of parity in odd dimensions
and the related Chern-Simons action [31,32]. It will be considered
in Section 7.

{\sl 5.2 Minimal form of the anomaly}\smallskip

The presence of the anomaly indicates that the effective action
has terms which are not chirally invariant. As it is well known
different regularizations in principle produce effective actions
which differ in local polynomial terms. Given the fact that such different
actions are related by counterterms, whose parameters are to be fixed
anyway by some renormalization prescription, they are considered
physically equivalent. These differences reflect in turn in the form
of the anomaly. The different regularizations give anomalies
differing by so called unessential terms, that is, terms which
can be obtained as the variation of local polynomial actions.
Bardeen [15] worked out the four dimensional case, including
vector, axial-vector, scalar and pseudoscalar fields in the Dirac operator.
He showed that the scalar and pseudoscalar fields did not
contribute to the essential anomaly and moreover that the only essential
terms were those of abnormal pseudoparity, i.e. containing the Levi-Civita
pseudotensor and thus purely imaginary in Euclidean space. This
Bardeen's or minimal anomaly were later shown to derive from the
Wess-Zumino-Witten action [17,20], which of course is not a local
polynomial.
Since the work of Bardeen, it has been shown that nongauge fields [33,34],
and internal gauge fields (i.e. transforming homogeneously under chiral
transformations) [35] do not contribute to the essential anomaly.
This is also suggested by the fact that if the formal variation of the action
is regularized in a chirally covariant way, there is an obstruction to the
integrability conditions which depends on the vector and axial gauge
fields only [2,12].

Let us restate this result with our formalism in a way which is easily
extended to higher dimensions and more general theories, i.e. relying only
on algebraic transformation properties but not on the detailed coupling
structure of the external fields. That is, let us show that in fact for
a completely general Dirac operator in two and four dimensions in Euclidean
flat space one can write enough counterterms to bring the
anomaly to Bardeen's form. To be precise, let $\D$ be
of the form $\D=\Dzero+\X$ where $\X$ is a local function, and
$\Dzero$ and $\X$ transform independently under axial transformations
$$\Dzero \underarrow{A} e^{-i\alpha\gamma_5}\Dzero e^{-i\alpha\gamma_5},
\qquad\X\underarrow{A} e^{-i\alpha\gamma_5}\X e^{-i\alpha\gamma_5}
\eqno(5.10)$$
$\Dzero$ contains the derivative part, and hence the
corresponding fields transform inhomogeneously. We want to show
that all the contributions to the anomaly coming from $\X$ can
be removed by counterterms. The construction is more easily
presented by following the approach of Ref. [35].
Let $\D_t = \Dzero + t\X$, then one can write the identity
$$W(\D) - W(\Dzero) = \int_0^1\dd t\,\delta_XW(\D_t)=
\int_0^1\dd t\int\dd^Dx\,\tr\left(\X \J(\D_t)\right)\eqno(5.11)$$
Here $\J(\D)$ is the current introduced in Eq. (2.11). The axial anomalous
contribution to the action containing $\X$ will be local polynomials
if and only if the current has the form
$$\J(\D) = \J_c(\D) + \P(\D) \eqno(5.12)$$
where $\P$ is a local polynomial and $\J_c$ is a chiral
covariant current, transforming as
$$\J_c \underarrow{A} e^{i\alpha\gamma_5}\J_c e^{i\alpha\gamma_5} \eqno(5.13)$$
so that $\delta_A\Tr(\X\J_c) = 0$. Notice the opposite sign for the
axial transformation as compared to the Dirac operator, Eq. (5.1).
Indeed, the above decomposition of the current yields an analogous
separation for the action, namely
$$W(\D) = W(\Dzero) + \int_0^1\dd t\int\dd^Dx\,\tr(\X\J_c(\D_t)) +
\int_0^1\dd t\int\dd^Dx\,\tr(\X\P(\D_t)) \eqno(5.14)$$
The first term gives ${\cal A}_A(\D_0)$, the second is chiral
invariant and the last is the local polynomial counterterm.
The observation that the
current is of the form (5.12) was already made in Ref. [26]
for chiral fermions in the presence of gauge fields. As already
pointed out by Bardeen and Zumino, the fact that the total
current admits such a decomposition is not obvious and requires a
constructive proof for each case.

Let us construct $\P$ explicitly for the two dimensional case. Clearly
$\P$ must satisfy the conditions
$$\delta_A\P = \delta_A\J = {\delta{\cal A}_A\over \delta\D}\,, \qquad
\delta_g\P=0 \eqno(5.15)$$
as a consequence of Eq. (5.12) and vector gauge invariance.
Notice that since the anomaly involves the symbol $\bra\ \ket$ and $\D$ is
not a multiplicative operator (see the remark to Eq. (4.16)),
cyclic property might not be applied
in principle to compute $\delta{\cal A}_A/\delta\D$. Nevertheless,
the vector gauge invariance of the anomaly and the fact that $\delta\D=\X$
is a multiplicative operator, allows to do so, yielding
$\delta_X{\cal A}_A=\bra\X\{2i\alpha\gamma_5,\D\}\ket$.
One can check that the first relation is then satisfied by
$$\P_0 = -{1\over 4\pi}\bra x|\D|0\ket  \eqno(5.16)$$
Unfortunately this solution breaks vector gauge invariance. In order to
construct $\P$ we should subtract a new polynomial $\P_1$ from $\P_0$
to reestablish vector gauge invariance. Also $\P_1$ must
be axial covariant in order not to modify the already correct axial
transformation of $\P_0$
$$\delta_g\P_1 = \delta_g\P_0\,, \quad\quad
\P_1\underarrow{A} e^{i\alpha\gamma_5}\P_1 e^{i\alpha\gamma_5} \eqno(5.17)$$
The only object algebraically made out of $\D$, transforming axially as
$\P_1$ should, is $\gamma_\mu\D\gamma_\mu$, which however is also
vector gauge invariant and hence cannot match $\delta_g\P_0$.
Therefore we must resort to new objects or use more information on $\D$ to
write enough counterterms. Let $\Db$ be other Dirac operator with
$$\delta_g\Db = \delta_g\D = \thru{a},
\quad\quad \Db \underarrow{A} e^{i\alpha\gamma_5}\Db e^{i\alpha\gamma_5}
\eqno(5.18)$$
If $\D$ has the standard hermiticity, (i.e. the Hamiltonian is Hermitian in
Minkowski space), a solution is provided by $\Db=-\D^\dagger$. Now it is
straightforward to obtain $\P_1$ by writing all the possible
objects of second order with the correct axial transformation,
constructed with $\D$, $\Db$ and $\gamma_\mu$ and adjusting
their coefficients to match $\delta_g\P_0$. We finally find the solution
$$\P = -{1\over 4\pi}\bra x|(\D-\Db)|0\ket \eqno(5.19)$$

In the previous construction $\P_0$ has been obtained by trial and error.
In more complicated cases, the best way to proceed is to introduce a
polynomial action in the Dirac operator, $W_0(\D)$,
from which $\P_0$ formally derives, since the number of possible terms
decreases substantially. The fact that the anomaly cannot be subtracted by
local and polynomial counterterms, prevents the existence of such an
action in a literal sense. Nevertheless one can impose
$\delta_XW_0(\D) =\int\dd^Dx\,\tr(\X\P_0(\D))$ and $\delta_AW_0(\D)={\cal A}_A$
modulo commutator terms, which would vanish if cyclic property
were valid, i.e. if all the operators involved were multiplicative.
That is what we mean by formal in this context.
There are two key observations. First, that
actions $W_0(\D)$ which are algebraically made out of $\D$, uniquely determine
the current $\P_0(\D)$, namely, by substituting $\D\to\D+\X$ in $W_0(\D)$,
keeping terms with just one $\X$ and freely using the cyclic property
to bring all the $\X$ say to the left. The relation is unique even if
$\P_0(\D)$ is only the formal variation of $W_0(\D)$.
And second, if the current $\P_0(\D)$ reproduces the variation of the axial
anomaly, i.e. $\delta_A\P_0(\D)=\delta{\cal A}_A/\delta\D$ (see Eq. (5.15)),
the action must reproduce the anomaly $\delta_AW_0(\D)={\cal A}_A$
at the formal level. It is therefore advantageous to solve this latter equation
and hence to derive $\P_0(\D)$.

One can check that a solution in the two-dimensional case, is
given by the action
$$ W_0(\D) = -\Bra\,{1 \over 2}\D^2\,\Ket =-{1\over 8\pi}\int \dd^2x
\,\tr\bra x|\D^2|0\ket \eqno(5.20)$$
which of course is not gauge invariant. $W_0(\D)$ formally gives the axial
anomaly, and its current is the same $\P_0(\D)$ found previously, Eq. (5.16).
In summary,
$\P_0$ is a local polynomial current with the same anomaly as $\J$.
Once $\P_0$ is available one can proceed as explained above to obtain $\P$.

The previous method can be applied to the four dimensional case.
We find for $W_0$ and $\P$ the following expressions
$$\eqalign{
W_0(\D) &= -\bra {1\over 24}\D^4
+{1\over 24}\D^2\gamma_\mu\D^2\gamma_\mu + {1\over 12}\D^3\gamma_\mu\D
\gamma_\mu\ket \cr
\P &= -{1\over 12}{1\over (4\pi)^2}\bra x|
(2(\D^3-\Db\D\Db) +\D\gamma_\mu\D^2\gamma_\mu
-\Db\gamma_\mu\Db\D\gamma_\mu \cr
 &+ \gamma_\mu\D^2\gamma_\mu\D -\gamma_\mu\D\Db\gamma_\mu\Db
 +\D\gamma_\mu\D\gamma_\mu\D -\Db\gamma_\mu\Db\gamma_\mu\Db \cr
&+\gamma_\mu\D^3\gamma_\mu
-\gamma_\mu\D\Db\D\gamma_\mu + \D^2\gamma_\mu\D\gamma_\mu
-\Db\D\gamma_\mu\D\gamma_\mu \cr
&+ \gamma_\mu\D\gamma_\mu\D^2-\gamma_\mu\D\gamma_\mu\D\Db )|0\ket\cr }
\eqno(5.21)$$
The terms without $\Db$ are those coming from $\P_0$. This polynomial
generalizes that found in Ref. [26].
This completes the proof that all the fields in $\D$ transforming
homogeneously under chiral rotations do not contribute to the essential
anomaly in two and four dimensions.
If one keeps only external vector and axial fields in
$\Dzero$, ${\cal A}_A(\Dzero)$ is Bardeen's anomaly (up to some
normal pseudoparity terms which can again be removed by counterterms).
Explicit expressions for the counterterms after having worked out
the Dirac algebra can be looked up at Appendix D.

An interesting aspect of the previous results is that the counterterms needed
to reproduce the Bardeen form of the anomaly in $\zeta$-function regularization
requires introducing, besides $\D$,
a new Dirac operator $\Db$ transforming in the same way as the current
under the chiral group.
As already mentioned, an operator transforming in that way in Euclidean
space is given by $-\D^\dagger$. This agrees with similar findings in
other regularization schemes, using ad hoc prescriptions like e.g. to
separate the action into real and imaginary parts $W(\D)=W^+(\D)+W^-(\D)$,
by means of the formula
$$\eqalign{
W^+(\D) &=
-{1\over 4}\left\{\Tr\log(\D\D^\dagger)+\Tr\log(\D^\dagger\D)\right\}\cr
W^-(\D) &= -{1\over 4}\left\{\Tr\log(\D^2)-\Tr\log(\D^{\dagger 2})\right\}
\cr}\eqno(5.22)$$
and similar ones [2,10,36,37].

In contrast to the axial anomaly, the scale anomaly
${\cal A}_S$ contains no unessential terms: by dimensional counting the
possible local polynomials
would be scale invariant or else would have to include external functions,
thus introducing an anomaly in the Poincar\'e symmetry.

\medskip
{\bf 6. Gradient expansion}
\medskip

{\sl 6.1 General considerations}\smallskip

In Section 4 we considered an inverse mass expansion for the effective
action of $\D$. It was both an expansion in the number of external
fields and the number of derivatives. Here we shall consider an
expansion in the number of derivatives and the number of fields with
Lorentz indices. That is, we take
$$\D = \M + \Dzero\,,\qquad \M(x) = S(x) +i\gamma_5 P(x)  \eqno(6.1)$$
where $S(x)$ and $P(x)$ are scalar and pseudoscalar fields and $\D_0$
includes $\ds$ as well as vector, axial vector, tensor fields, etc, and
we expand in powers of $\D_0$. This is a resummation of the inverse
mass expansion, relaxing the restriction that $S$ and $P$ should be
weak fields. Another standard resummation, complementary to this one,
is the perturbative expansion which assumes weak but not necessarily
smooth fields.
The gradient expansion is a semiclassical expansion similar to that
used in quantum mechanics and many body physics [38,39]. This means that
its starting point approximates the spectrum of $\D$ by a continuum.
Discretization effects are averaged and cannot be recovered in detail
by resummation, hence the expansion is at most only asymptotic.
The same is true of course for the inverse mass expansion.

Expanding eq. (3.6) and using (3.7), we have
$$ \eqalign{
\Tr\D^s &=
\sum_{N=0}^\infty\left[\Tr\D^s\right]_N \cr
\left[\Tr\D^s\right]_N &= (-1)^{N+1}
\int{\dd^Dx\dd^Dp\over
(2\pi)^D}\int_\Gamma{\dd z\over 2\pi i}
z^s\tr\bra x|(G_0\D_0)^NG_0|0\ket \cr} \eqno(6.2)$$
$$ \eqalign{
G_0(x,p;z) &= (\thru{p}-z+\M)^{-1} \cr
 &=-(p^2+(z-S)^2+P^2+i\gamma_5[S,P])^{-1}(\thru{p}+z-S+i\gamma_5P)\cr}
\eqno(6.3)$$
Let us remark that for $N>D$ the integrals are ultraviolet finite, yet we
cannot proceed formally by simply taking
$\bra\log(\M+\thru{p}+\Dzero)\ket$ and expanding
in powers of $\Dzero$ because this would require trace cyclicity and in fact
gives wrong results (also we cannot expand formally $\Tr\log(\M+\Dzero)$
before using the Wigner transformation method because every term would
diverge). However it can be done for the current or,
equivalently, we can take $s=-1$ above and perform
the $z$ integral first. Thus, we obtain for the current in $N$-th order, with
$N>D$
$$[J(x)]_N = (-1)^{N+1}\int{\dd^Dp\over(2\pi)^D}
\tr\bra x|\left({1\over\M+\thru{p}}
\Dzero\right)^N{1\over\M+\thru{p}}|0\ket\,,\qquad N>D
\eqno(6.4)$$
The action can then be reconstructed with the Eq. (2.14).
Because $z$ no longer
appears in Eq. (6.4), $\J_{N>D}$ enjoys all the classical symmetries of
$\D$; so long as the symmetry transformation
does not mix different orders of the expansion, which is the case
for the usual Poincar\'e invariant internal symmetries.

If $\D_0$ contains only fields with an odd number of Lorentz
indices, the terms with $N$ odd vanish after performing the $p$\ integration
in even dimension.

\medskip
{\sl 6.2 Effective action in 1+1 dimensions}\smallskip

The problem with gradient expansions is that it is in
general difficult to
work out explicitly the inversion of matrices implied in $G_0(x,p;z)$, which
is necessary to perform the $z$ and $p$ integrals [9].
To be concrete and keep the computations simple, we will consider the
1+1 dimensional case with SU(2) flavor symmetry, that is
$$\D =\ds+\thruv{V} + \thru{A}\gamma_5 +S+ i\gamma_5P \eqno(6.5)$$
where $S(x)$ is a Hermitian c-number and $P(x)$, $V_\mu(x)$ and
$A_\mu(x)$ are matrices in the fundamental representation of su(2).
Because of the two
dimensional identity $\gamma_\mu\gamma_5 = i\epsilon_{\mu\nu}\gamma_\nu\,,$
we can simplify the amount of algebra without loss of generality
by reabsorbing $A_\mu(x)$ in $V_\mu(x)$  with
$\tilde{V}_\mu=V_\mu-i\epsilon_{\mu\nu}A_\nu$.

Let ${\cal L}$ be the matrices in the fundamental representation of sl(2,C),
i.e. of the form $a_i\tau_i$ where $\tau_i$ are the Pauli matrices and
$a_i$ are c-numbers.
The following properties of ${\cal L}$ will be widely used later:
$$ \hbox{\rm for } A,B\in {\cal L},\qquad \tr A=0,\quad [A,B]\in
{\cal L},\quad \{A,B\} = \hbox{\rm c-number} \eqno(6.6)$$
In particular $G_0(x,p;z)$ can be computed to obtain
$$\left[\Tr\D^s\right]_N =
\int{\dd^2x\dd^2p\over (2\pi)^2}\int_\Gamma{\dd z\over 2\pi i} z^s
\tr\bra x|\left(\left({\thru{p}+\sigma+i\gamma_5P\over\Delta}
\Dcst\right)^N {\thru{p}+\sigma+i\gamma_5P\over\Delta}\right)|0\ket
\eqno(6.7)$$
where we have defined
$$\sigma=z-S\qquad\Delta=p^2+\sigma^2+P^2\,,\qquad \Dct_\mu=i\partial_\mu+
\tilde{V}_\mu \eqno(6.8)$$
and $\sigma$, $P^2$ and $\Delta$ are c-numbers. Odd orders vanish.
Let us compute explicitly the first two terms, $N=0,2$.
One finds integrals of the following form
$$I_2(k,N,D,s;S,P) = \int{\dd^Dp\over(2\pi)^D}\int_\Gamma{\dd z\over
2\pi i} { z^sp^{2k}\over \Delta^N} \eqno(6.9)$$
Because the integral is a meromorphic function of $D$, we can
perform first the integral in $p$ as in dimensional
regularization, and then the $z$ integral gives the
hypergeometric function ${}_2F_1$ [40, p. 555]:
$$\eqalign{
I_2 =-{1\over (4\pi)^{D/2}} & {\Gamma(N-n)\Gamma(2n-s-1)\over
\Gamma({D\over 2})\Gamma(N)\Gamma(n-s)}
\times\left(
e^{+i\pi n}z_+^{s-2n+1}\,{}_2F_1(n,2n-s-1;n-s;{z_-\over z_+})
 \right. \cr & + \left.
e^{-i\pi n}z_-^{s-2n+1}\,{}_2F_1(n,2n-s-1;n-s;{z_+\over z_-})
\right) \cr} \eqno(6.10)$$
with $n=N-{1\over 2}D-k$, \  $z_\pm=S\pm i|P|$ and $|P|=(P^2)^{1\over 2}$.
For even $D$ it is convenient to write the same result as
$$I_2={1\over (4\pi)^{D/2}}
{\Gamma(N-n)\over\Gamma({D\over 2})\Gamma(N)} \sum_{z=z_\pm}\left({\partial
\over\partial z}\right)^{n-1} {z^s\over(z-z^*)^n}\eqno(6.11)$$
with the prescription
$$\hbox{\rm for }\alpha,\beta\in\hbox{\rm C},\qquad
\left({\partial\over\partial z}\right)^\alpha z^\beta = {\Gamma(\beta+1)\over
\Gamma(\beta-\alpha+1)}z^{\beta-\alpha}\,,\qquad
\left({\partial\over\partial z}\right)^\alpha {z^*}^\beta =0
\eqno(6.12)$$
In this form eq. (6.11) holds too if $z^s$ is replaced by any
function analytic in C$-$R${}^-$ and sufficiently convergent
at $\infty$.
After taking the trace in Dirac space we have
$$\left[\Tr\D^s\right]_{N=0}
= 2\Bra\sum_{z=z_\pm}\left({\partial\over\partial z}\right)^{-1}
z^s(z-S)\Ket
= 4\,\Re\Bra{z_+^{s+2}\over s+2}-S{z_+^{s+1}\over s+1}\Ket\,,
\quad s\in{\rm R} \eqno(6.13)$$
And hence for the action at zeroth order
$$W_0 =\Bra M^2\left(\log({M^2\over\mu^2})-3\right) +
2P^2\Ket\eqno(6.14)$$
$\bra\ \ket$ was defined in Eq. (4.15) and in these formulas $\tr$ acts on
flavor space only, and $M^2=S^2+P^2$. A similar calculation in
four dimensions gives
$$W_0 = -\Bra M^4\left(\log({M^2\over\mu^2})-{25\over6}\right)
+{4\over 3}M^2P^2 + {4\over 3}P^4\Ket \eqno(6.15)$$
The polynomial chiral breaking terms yield the spurious axial
anomaly discussed in the previous section.

Let us compute the second order term. After taking the
trace in Dirac space, we can always proceed by carrying
the covariant derivatives $\Dct_\mu$ to the right, for instance
$$\Dct_\mu \Delta^{-1} = -\Delta^{-2} (-2\sigma[\Dct_\mu,S] +
\{P,[\Dct_\mu,P]\}) + \Delta^{-1}\Dct_\mu \eqno(6.16)$$
so that the various $\Delta$ and $\sigma$ are collected to the
left and the integral $I_2$ applies. This is not the most
efficient strategy in this case but it is systematic and allows
for algebraic manipulator implementation in more complicated cases.
Due to vector covariance, one finds that $\Dct_\mu$ appears only
covariantly, that is in the form $[\Dct_\mu,X]$. The vanishing
of the non covariant terms follows after the $p$ integration
and using the ${\cal L}$ identities, eq. (6.6).
The result can be brought to a more symmetric form by using
integration by parts of the covariant derivatives and trace cyclicity:
$$\eqalign{
W_2 &= \Bra -{1\over 12}{(\tilde{D}_\mu M^2)^2 \over M^4}+
{1\over 2}{(\tilde{D}_\mu S)^2 + (\tilde{D}_\mu P)^2\over M^2}\right. \cr
 &\left.+\arctan({P\over S}) i\epsilon_{\mu\nu}\tilde{F}_{\mu\nu}
+{1\over 2}({1\over P^2} {S\over M^2} -{1\over P^3}\arctan({P\over S}))
\epsilon_{\mu\nu}P\tilde{D}_\mu P\tilde{D}_\nu P \Ket\cr}
\eqno(6.17)$$
Here $\tilde{D}_\mu X=[\tilde{\cal D}_\mu,X]$,
$i\tilde{F}_{\mu\nu}=[\Dct_\mu,\Dct_\nu]$.

\medskip
{\sl 6.3 Analysis of the effective action. Wess-Zumino-Witten action}
\smallskip

In order to obtain a separation $W_2 = W_2^+ + W_2^-$ in two terms with well
defined hermiticity and pseudoparity, we have to reexpand
$\Dct_\mu$ as $\tilde{D}_\mu X=D_\mu X-\epsilon_{\mu\nu}[A_\nu,X]$.
The real part, $W_2^+$, comes from terms without explicit $\epsilon_{\mu\nu}$
and even number of axial fields or else with explicit $\epsilon_{\mu\nu}$
and odd number of axial fields, whereas $W_2^-$ is the other
way around.
In principle, the effective action up to second order in the gradient
expansion should saturate the anomaly equations. This can be trivially
checked for the scale anomaly. Although the resulting expressions
are manifestly vector gauge invariant, the corresponding check for the
axial anomaly requires more work. A convenient method to do so is
to explicitly establish that the anomalous terms of the computed effective
action coincide with the Wess-Zumino-Witten action plus additional
polynomial counterterms.

To analyze the chiral transformation properties of
$W_2$, let $M(x)$, $U(x)$ and $\phi(x)$ be defined by
$$S+iP=MU,\quad\qquad U=e^{i\phi}=\cos\phi+i\sin\phi \eqno(6.18)$$
hence $U\in$ SU(2),\ \  $\phi,\sin\phi\in$ su(2) and
$M$, $\cos\phi$ are c-numbers. $\phi(x)$ is a multivalued
function of $S$ and $P$, well defined everywhere by assuming
explicitly that $M(x)>0$, or equivalently that $\M$
is non singular, as required by Seeley's formula (2.7).

Under the previous assumptions, the terms in $W_2$ with
explicit $\epsilon_{\mu\nu}$ turn out not to depend on $M(x)$,
as it is readily shown. That is, this part of $W_2$ can be written as
$$W_2^{\epsilon_{\mu\nu}}=\Bra 2i\tilde{F}\varphi + {1\over 4}
{\sin(2\phi)-2\varphi\over\sin^3\phi}\sin\phi(\tilde{D}\sin\phi)^2\Ket
\eqno(6.19)$$
$$\varphi=\phi +n\pi{\phi\over|\phi|}\,,\qquad n\in{\rm Z}\eqno(6.20)$$
$\varphi$ is such that $\tan\varphi=\tan\phi$ and takes into account
explicitly the multivaluation. A standard exterior algebra
notation has been used [26],
$$\tilde{F}={1\over 2}F_{\mu\nu}\dd x_\mu\dd x_\nu\,,\qquad
\tilde{D} X=\tilde{D}_\mu X\dd x_\mu \eqno(6.21)$$
and $\dd x_\mu$ are anticommuting variables.
For more general symmetry groups, $M(x)$, defined by the
factorization $S+iP=MU$ with $M$ Hermitian and $U$ unitary, will not be
a c-number and will not cancel in $W^-$. Nevertheless, the symmetry under
local rescaling $\M(x)\to\lambda(x)\M(x)$, where $\lambda(x)$ is a c-number
local function, still holds. This general fact will be established below
when dealing with the fermionic current.

In order to work out the expressions it is convenient to use the
following identities, particularly useful for computing derivatives
of the fields:
$$\hbox{\rm for }a\in{\cal L},\quad
\delta f_+(a)={f^\prime_+(a)\over 2a}\{a,\delta a\},\quad
\delta f_-(a)={f^\prime_-(a)\over 2a}\{a,\delta a\}
+{f_-(a)\over 2 a^2}[a,\delta a] \eqno(6.22)$$
where $f_{\pm}$ are arbitrary even and odd functions respectively,
and $f^\prime_{\pm}$ their derivatives. As a consequence
$$\eqalign{
iU^\dagger\delta U &= -{1\over\cos\phi}\delta\sin\phi + {1\over 2}
(\tan\phi+i)[\sin\phi,\delta\sin\phi] \cr
iU\delta U^\dagger &= +{1\over\cos\phi}\delta\sin\phi - {1\over 2}
(\tan\phi - i)[\sin\phi,\delta\sin\phi] \cr }\eqno(6.23)$$
After some algebra, the real part of $W_2$ can be written as
$$W_2^+=\Bra{1\over 24}\left({\partial_\mu M^2\over
M^2}\right)^2 +{1\over 2}{\cal R}_\mu^2 -2 A_\mu^2 \Ket \eqno(6.24)$$
where we have defined
$${\cal R}_\mu = U^\dagger i\partial_\mu U - V^R_\mu + U^\dagger V^L_\mu U\,,
\quad\quad V^{R,L}_\mu = V_\mu\pm A_\mu \eqno(6.25)$$
${\cal R}_\mu$ transforms covariantly under chiral gauge transformations
$$\Omega_{R,L}(x)\in\hbox{SU(2)},\qquad
U\to\Omega_LU\Omega_R^\dagger\,,\qquad V^{R,L}_\mu\to\Omega_{R,L}
(i\partial_\mu+V_\mu^{R,L})\Omega_{R,L}^\dagger \eqno(6.26)$$
Vector gauge transformations correspond to the diagonal
subgroup. $M$ is chiral invariant and the only chiral breaking
term in $W_2^+$ is $A_\mu^2$, which is a polynomial.

Similarly, the imaginary part can be brought to the following form
$$\eqalign{
W_2^- &= \Gamma_{\rm WZW}(U) + \Gamma_G(U,V^R,V^L)-\Gamma_G(I,V^R,V^L) \cr
\Gamma_{\rm WZW}(U) &= {1\over 12\pi}\int_{D_3}\tr(U^\dagger i\dd U)^3 \cr
\Gamma_G(U,V^R,V^L) &= \Bra V^RU^\dagger\dd U-V^LU\dd U^\dagger
-i V^RU^\dagger V^L U\Ket \cr} \eqno(6.27) $$
where $\dd X = \partial_\mu X\dd x_\mu$. $\Gamma_{\rm WZW}$ is the
correctly normalized Wess-Zumino-Witten action in two dimensions.
$\Gamma_G(U,V^R,V^L)$ is obtained by
chirally gauging $\Gamma_{\rm WZW}$ [20]. The last term is known
as Bardeen's subtraction, a counterterm which reestablishes
vector gauge invariance. There $I$ stands for the identity of SU(2).

The way $\Gamma_{\rm WZW}$ appears deserves some comment. The field
configuration
$U(x)$ defines a two dimensional manifold $M_2$ without
boundary inside SU(2), and $D_3$ is a three dimensional manifold such that
$\partial D_3=M_2$. Up to a quantized multivaluation [20],
$\Gamma_{\rm WZW}$ does not depend on the choice of $D_3$ because
$\omega_3={1\over 12\pi} \tr(U^\dagger i\dd U)^3$ is closed.
Hence, there is a 2-form $\Omega_2$ such that locally $\omega_3=\dd\Omega_2$.
However $\Omega_2$ cannot be regular everywhere, because
$\omega_3$ is not exact (in fact it is just the volume element in SU(2)).
Choosing $D_3$ to avoid the Dirac singularity one has
$$\Gamma_{\rm WZW} = \int_{D_3}\omega_3 = \int_{M_2}\Omega_2
\eqno(6.28)$$
To arrive to eq. (6.27), what one can show is that a valid
choice for $\Omega_2$ is
$$ \Omega_2 = {1\over 16\pi} \tr\left( {\sin(2\phi) -2\varphi \over
\sin^3\phi}\sin\phi(\dd\sin\phi)^2\right) \eqno(6.29)$$
which is just $W_2^{\epsilon_{\mu\nu}}$ in eq. (6.19) for
$V_\mu=A_\mu=0$, and arbitrary $n$. The singularity is at
$U=(-I)^{n+1}$. In fact the solution of $\omega_3=\dd\Omega_2$ is
unique imposing global vector invariance and regularity at
$\phi=0$. Global chiral invariance is not manifest for any
choice of $\Omega_2$. Also one can check that by varying $n$,
$\int\Omega_2$ changes by an integer multiple of $2\pi i$.
Another interesting point is that vector gauge invariance would be
achieved by minimal coupling $i\dd\to i\dd+\tilde{V}$ in
$\Omega_2$. However such an action would not be single valued
modulo $2\pi i$. The new vector gauge invariant term
$\bra 2i\tilde{F}\varphi\ket$ in $W_2^{\epsilon_{\mu\nu}}$
reestablishes the one-valuedness of the action.

$W_2^-$ can be written in other interesting form by applying Stokes'
theorem to $\Gamma_G$, namely
$$W_2^- = {1\over 4\pi}\int_{D_3}\tr({1\over 6}{\cal R}^3
-i{\cal R}{\cal F}_R -2iA{\cal F}_R +{4\over 3}A^3) -\hbox{p.c.}
\eqno(6.30)$$
where ${\cal R}=U^\dagger i\dd U -V_R+U^\dagger V_L U$, ${\cal F}_R
=\dd V_R-iV_R^2$ are chiral covariant, and p.c. means to subtract
the parity conjugate terms, i.e. the same terms exchanging
$U\leftrightarrow U^\dagger$,
$V_R\leftrightarrow V_L$ and $A\leftrightarrow -A$. This form is manifestly
vector gauge invariant and the axial anomalous terms (those with $A_\mu$)
are independent of $U$. These terms are polynomial, yet they cannot be
removed by counterterms because they do not form a closed 3-form by
themselves.

We would like to emphasize that ours is an ab initio calculation
of the action. Once the $\zeta$-function prescription is adopted,
there is no more freedom nor ambiguity in the calculation. This
is in contrast to derivations of $\Gamma_{\rm WZW}$ by
integration of the chiral anomaly, which have to assume that
$S+iP$ lies on the chiral circle, i.e. $M(x)$ constant [41-44].

\medskip
{\sl 6.4 The fermionic number current}
\smallskip

As another illustration of the Wigner transformation technique, let us consider
the fermionic number current $-\psi\gamma_\mu\psi$ in four dimensions with
arbitrary internal GL(n,C) symmetry.
It is obtained from
$$\delta_\omega\D = -\thru\omega(x),\qquad
\delta_\omega W = \int\dd^Dx\,\omega_\mu(x)j_\mu(x) \eqno(6.31)$$
where $\omega_\mu(x)$ is a c-number field. As an example, for
$\D= \ds + \thruv{V}(x)+m$, a simple calculation from Eqs. (4.7) and
(4.17) gives
$$W = -{2\over 3}\bra F_{\mu\nu}^2\ket \log\left({m\over\mu}\right) +
O(m^2) \eqno(6.32)$$
which gives a contribution to $j_\mu(x)$. Here we will consider the anomalous
part of the current, $j_\mu^-(x)$ in the presence of scalar and pseudo scalar
gl(n,C)-matrix valued fields
$$\D = \ds +\M(x)\,,\qquad \M(x)=S(x)+i\gamma_5 P(x)\eqno(6.33)$$
Because the pseudoparity odd (p.o.) part of the current
is ultraviolet finite we
will use directly the expression
$$\delta_\omega W^- = \Tr\left(\thru{\omega}{1\over \D}
\right)_{\rm p.o.}
\eqno(6.34)$$
On the other hand, the lowest order contribution appears at fourth order in a
gradient expansion of $W$, due to the presence of the Levi-Civita pseudotensor.
Thus to lowest order, the anomalous current is obtained from
$$\delta_\omega W^-_{\rm lowest} =
-\int{\dd^4x\dd^4p\over(2\pi)^4}\tr\bra x|\thru{\omega}{1\over \thru{p}+\M}
\left(\ds{1\over \thru{p}+\M}\right)^3|0\ket_{\rm p.o.}\eqno(6.35)$$
It is convenient to introduce the definitions
$$\eqalign{
M_{R,L} &= S\pm iP\,,\qquad P_{R,L}= {1\over 2}(1\pm\gamma_5) \cr
G_{LR} &= (p^2+ M_L M_R) ^{-1}\,,\qquad G_{RL} =(p^2+M_R M_L)^{-1} \cr}
\eqno(6.36)$$
so that
$$\eqalign{
(\thru{p}+\M)^{-1}&= -(p^2+\M^\dagger\M)^{-1}(\thru{p}-\M^\dagger)=\cr
 &= P_RG_{LR}M_LP_R + P_LG_{RL}M_RP_L
-P_RG_{LR}\thru{p}P_L -P_LG_{RL}\thru{p}P_R \cr}\eqno(6.37)$$
Substituting in $\delta_\omega W^-$, and keeping only terms with
$\gamma_5$, one obtains
$$\eqalign{
\delta_\omega W^-_{\rm lowest} &=
-2i\int{\dd^4p\over(2\pi)^4}\omega\tr\big(
G_{LR}M_L\dd G_{RL}M_R\dd G_{LR}M_L\dd G_{RL}M_R\cr
&+ {1\over 2}p^2(G_{LR}\dd G_{LR}\dd G_{LR}M_L\dd G_{RL}M_R-
G_{LR}\dd G_{LR}M_L\dd G_{RL}M_R\dd G_{LR})\cr
&+ {1\over 2}p^2 (G_{LR}M_L\dd G_{RL}\dd G_{RL}\dd G_{RL}M_R +
G_{LR}M_L\dd G_{RL}M_R\dd G_{LR}\dd G_{LR})\big) - {\rm p.c.} \cr
}\eqno(6.38)$$
The parity conjugate (p.c.) is obtained by exchanging the labels $R$
and $L$ everywhere, tr no longer includes Dirac trace, and
an exterior algebra notation has been used with
$\omega = \omega_\mu\dd x_\mu$ and $\dd=\dd x_\mu\partial_\mu$.
$\dd$ derivates to
the right until it finds another $\dd$ or gives zero if it reaches the
right end.
The integral over $p$ is convergent, but it cannot be done in closed form for
arbitrary $M_{R,L}$ fields.

An interesting property of $\delta_\omega W^-$, which can already be derived
without explicitly performing the momentum integral, is its invariance
under the local rescaling $S(x)\to \lambda(x)S(x)$ and $P(x)\to\lambda(x)P(x)$,
with $\lambda(x)$ an arbitrary local c-number function. Due to the close
relation between the fermionic number current and the imaginary part of the
effective action [45,46], to be addressed below, this property holds
for $W^-$ at lowest nonvanishing order as well.
As a consequence, in the particular case of $M_R(x)=M(x)U(x)$,
$M(x)$ being a Hermitian c-number, not
necessarily constant, the expression simplifies:
$G_{LR}=G_{RL}=(p^2+M^2)^{-1}$ is also a c-number, all the terms
containing $\dd M(x)$ are readily shown to cancel,
the terms with $p^2$ vanish and the integral over $p$ is immediate,
with the following result
$$\delta_\omega W^-_{\rm lowest} =
-{1\over 24\pi^2}\int\omega\tr(R^3)\,,\qquad R= U^\dagger i\dd U
\eqno(6.39)$$
This gives the correctly normalized Goldstone-Wilczek current [19]
$$j_{\mu,\rm lowest}^-(x)=
-{1\over 24\pi^2} \epsilon_{\mu\nu\rho\sigma}\tr(R_\nu R_\rho
R_\sigma)\,,\qquad R_\mu=U^\dagger i\partial_\mu U \eqno(6.40)$$

Comparing Eqs. (6.27) and (6.40) we find that the 3-form $\omega_3$
is the integrand of the action in two dimensions and also it is the fermionic
number density in four dimensions, in both cases at lowest order and in
the pseudoparity odd sector.

\medskip
{\sl 6.5 Effective action from the fermionic number}\smallskip

The relation pointed out in the last paragraph of the previous subsection
is much more general as noted by several authors [45,46].
We will show subsequently that our formalism can handle the proof of
this general relation quite naturally. The general statement
is as follows. Let $\D$
be a Dirac operator in $D$ dimensions and $W^-$ the pseudoparity odd part
of its action. Consider the extension of $\D$ to a family of Dirac operators
$\D(v)$, $v\in[0,T]$ which interpolates between $\D$ at $v=T$ and
$\Dzero$ at $v=0$, where $\Dzero$ will be characterized below. Similarly
consider the extension of the space-time with two more dimensions $(u,v)$,
$u$ playing the role of Euclidean time, and of $\D(v)$ to a $D+2$ dimensional
Dirac operator $\hat\D^\prime$:
$$\eqalign{
&\hat\D^\prime = \hat\gamma_ui\partial_u + \hat\gamma_vi\partial_v+
\hat\D\cr
&\hat\gamma_u = i I\times\sigma_1\,,\quad\hat\gamma_v=iI\times\sigma_2\,,
\quad\hat\gamma_\mu=\gamma_\mu\times\sigma_3\,, \quad
\hat\gamma_5=\gamma_5\times\sigma_3 \cr}\eqno(6.41)$$
The $D+2$ matrices $\hat\gamma_a$ satisfy the Dirac algebra
$\{\hat\gamma_a,\hat\gamma_b\}=-2\delta_{ab}$, $a,b=u,v,0,\dots,D-1$, and
$\hat\gamma_5=i\hat\gamma_u\hat\gamma_v\gamma_5$ for even $D$.
The operator $\hat\D$ is defined by making the substitutions
$\gamma_{\mu,5}\to\hat\gamma_{\mu,5}$ in $\D(v)$. Thus it has the form
$$\hat\D =\left(\matrix{\D(v) & 0\cr 0 & \D^\dagger(v)\cr}\right)
 \eqno(6.42)$$
where $\D^\dagger$ appears because the hermiticity of the
fields in $\D$ is such that the Hamiltonian
$\hat\gamma_u(\hat\gamma_vi\partial_v + \hat\D)$
of the $D+2$ dimensional extended Dirac operator, is
Hermitian in Minkowski metric, and hence $\hat\D^\prime$ has the
standard hermiticity in Euclidean space.

The theorem establishing the relation between the fermion number and
the effective action can be stated in the following manner.
Let $B$ be the $D+1$-dimensional spatial region
defined by $(v,x_\mu)$, $v\in[0,T]$ and
$x_\mu\in{\rm R}^D$, and $F$ the fermion number enclosed in $B$ in the
$D+2$ dimensional system. Then for a suitable choice of the interpolating
path $\D(v)$, to be considered below, $2\pi F=-W^-$ in Euclidean metric,
$W^-$ being the $D$-dimensional pseudoparity odd effective action.
In fact, one can state this relation in a more general form, namely
$$\int_B\dd^Dx\dd v\,\tr\bra u,v,x|\hat\gamma_uf^\prime(\hat\D^\prime)
|u,v,x\ket =
{1\over 4 \pi}\left(\Tr f(\D)-\Tr f(\D^\dagger)\right)\big|_{v=0}^T
\eqno(6.43)$$
for a sufficiently convergent function $f(z)$ analytic in
C$-$R${}^-$, and
$f^\prime(z)$ its derivative. The relation between $F$ and $W^-$ is obtained
with $f(z) = \log z$ defined as the analytical extrapolation
of $\dd z^s/\dd s$ in $s$ at $s=0$.

Let us consider under which conditions this relation will be valid.
The contribution from the value $v=0$ to the r.h.s. of Eq. (6.43)
will vanish if $\Dzero$ is related
to $\Dzero^\dagger$ by a similarity transformation. This can always
be achieved for even $D$ by taking $\Dzero$ without pseudotensor fields,
i.e. without explicit $\gamma_5$,
since in this case $\Dzero^\dagger =\gamma_5\Dzero\gamma_5$.
Observe that $\hat\D^\prime$ is time independent
and does not contain fields with Lorentz index $u$, hence $u$ in the l.h.s.
must come from the Levi-Civita pseudotensor and it is
automatically pseudoparity odd as the r.h.s. Regarding the choice of the
interpolating path $\D(v)$, note that the l.h.s. can be
obtained from
$$\delta_\omega\hat\D^\prime = -\thru\omega\,,
\qquad\delta_\omega\Tr f(\hat\D^\prime) =
\int\dd^Dx\,\dd u\,\dd v \,\omega_a J_a \eqno(6.44)$$
Where $\thru\omega=\gamma_u\omega_u+\gamma_v\omega_v+\gamma_\mu\omega_\mu$
and $\mu=1,\dots,D$.
Because of vector gauge invariance, $\omega_a$ can only appear in the form
$\partial_a\omega_b-\partial_b\omega_a$, hence locally
$J_a = \partial_b J_{ab}$ with $J_{ab}$ antisymmetric.
Note that the l.h.s. of Eq. (6.43) can also
be written as $-\int_B\dd^Dx\dd v J_u = -\int_B\dd^Dx\dd v (\partial_v J_{uv}
+\partial_\mu J_{u\mu})$. The terms containing $J_{u\mu}$ cancel because
we are assuming $\D$ boundaryless in R$^D$.
The non vanishing contribution comes from $J_{uv}$ at the boundary, i.e.
$v=0$ and $v=T$. However, for even dimensions the $v=0$ contribution vanishes
since $J_{uv}$ must contain a $D+2$-dimensional Levi-Civita pseudotensor,
which cannot occur if $\Dzero$ is chosen as explained above.
If we take an adiabatic path, that is, a smooth path with $\partial_v\D(v)=
O(1/T)$ and $T\to \infty$, $J_{uv}|_{v=T}$ will depend only on $\D$ and $f(z)$
and not on the particular adiabatic path chosen.
For arbitrary paths $J_{uv}$ at $v=T$ would also depend on
$\partial_v^n\D|_{v=T}$, $n>0$, hence the result would be path dependent.
Let us remark that at lowest order $J_{ab}$ exists only locally if $f(z)$
is a multivalued
function such as $\log z$. In this case different homotopy classes of
adiabatic paths pick different branches of $f(z)$. An example is again
provided by
$\omega_3$ which is closed and hence locally exact. Indeed one of the
interesting aspects of the relation $2\pi F = -W^-$ is that $F$ is computed
as the integral of a single valued expression.

To proof Eq. (6.43) let us rewrite the l.h.s. using Seeley's representation
and the Wigner transformation trick in the space spanned by $|u,v\ket$
$$-\Tr\int_0^T\!\dd v\!\int\!{\dd^2\omega\over (2\pi)^2}\int_\Gamma{\dd z\over
2\pi i}f^\prime(z)\bra u,v|\hat\gamma_u{1\over
\hat\gamma_u(i\partial_u+\omega_u)+\gamma_v(i\partial_v+\omega_v)
+\hat\D -z}|0\ket \eqno(6.45)$$
$|0\ket$ is the state with zero momentum in the space $|u,v\ket$,
i.e. $\bra u,v|0\ket =1$, and $\partial_u$ and $\partial_v$ act to the
right until annihilating $|0\ket$. $\Tr$ includes trace in the space
$|x_\mu\ket$ as well as spinor and internal degrees of freedom in $D+2$
dimensions. The ultraviolet limit corresponds to integrate over
$\omega$ which is finite after the $z$ integration.
To proceed let us introduce the definitions
$$\eqalign{
& \Gamma = i\hat\gamma_u\hat\gamma_v=I\times\sigma_3\,,
\quad\Gamma=\Gamma^\dagger\,,\quad\Gamma^2=I \cr
& iH=\hat\gamma_u(\hat\D-z)\,,\qquad\{H,\Gamma\}=0 \cr}\eqno(6.46)$$
Manipulating the denominator of Eq. (6.45) as in (6.37) the formula can
be written as
$$\eqalign{
 -\Tr & \int_0^T\!\dd v\!\int\!{\dd^2\omega\over (2\pi)^2}
\int_\Gamma{\dd z\over 2\pi i}f^\prime(z)  \cr
&\times\bra u,v|{1\over (i\partial_u+\omega_u)^2+(i\partial_v+\omega_v)^2+H^2-
i\Gamma(\partial_vH)}
 (i\partial_u +\omega_u +\Gamma\partial_v-i\Gamma\omega_v +iH)
|0\ket \cr} \eqno(6.47)$$
We can set $\partial_u=0$ everywhere and also $\partial_v|0\ket=0$.
Furthermore in an expansion in powers of $\partial_v$, one can check
that only odd powers survive after doing an angular average over $\omega$
and taking trace using that $H$ is off-diagonal.
In particular for an adiabatic interpolating path, only the term linear in
$\partial_v$ survives. Then the expression takes the form
$$
-\Tr\int_0^T\!\dd v\!\int\!{\dd^2\omega\over (2\pi)^2}\int_\Gamma{\dd z\over
2\pi i}f^\prime(z)  \Gamma {1\over \omega^2+H^2}\left(
-(\partial_v{\omega^2\over \omega^2+H^2})-
(\partial_vH)H{1\over \omega^2+H^2}\right)
 \eqno(6.48)$$
Now we have the trace of an operator in the space $|x_\mu\ket$ only.
Because $f(z)$ is convergent the trace exists and
the cyclic property can be freely used in the two terms in Eq. (6.48).
The first term can be put in the form
$\partial_v(\Gamma g(H^2,\omega^2))$, and it vanishes after taking the trace
because $H^2$ is block diagonal with blocks $(\D^\dagger-z)(\D-z)$ and
$(\D-z)(\D^\dagger-z)$ which have the same spectrum. In the second term
we can integrate $\omega$, thus
$$\eqalign{
 -{1\over 4\pi}\Tr\int_0^T\!\dd v\!\int_\Gamma &
{\dd z\over 2\pi i}f^\prime(z)
 \Gamma\partial_v\hat\D{1\over \hat\D-z}
= {1\over 4\pi}\int_0^T\dd v\partial_v\Tr(\Gamma f(\hat\D)) \cr
&= {1\over 4\pi}\Tr(\Gamma f(\hat\D))\big|_{v=0}^T \cr}\eqno(6.49)$$
which coincides with the r.h.s of Eq. (6.43). This completes the proof.
Let us remark that the staircase relation (6.43) is valid beyond a gradient
expansion (in $D$ dimensions), and in case this expansion applies the leading
terms provide a relationship between a topological action and a topological
fermion current. Furthermore Eq. (6.43) does not rely on
the chiral circle constraint and holds for any space time dimension.

A rather amusing illustration of the relation comes from the zero dimensional
problem. In $D=0$ the most general Dirac operator is $\D=S+i\gamma_5 P$,
$S$, $P$ being Hermitian matrices in flavor space and $\gamma_5$
a $1\times 1$ matrix which equals $-1$ in our convention (see Eq. (2.2)).
Certainly, the Grassmann
integral (2.5) can be carried out and $W=-\tr\log(\D/\mu)$ exactly, without
regularization. Remarkably, the effective action is ultraviolet finite
yet it possesses both scale anomaly and (essential) axial anomaly
(even if there are no gauge fields). This is also in agreement with the general
formulas (5.3) and (5.8).

In the Abelian case $\D=Me^{i\gamma_5\phi}$,
$M,\phi\in$R, the action is simply $W=-\log(M/\mu)+i\phi$.
On the other hand, the pseudoparity odd singlet current in two dimensions
($0+2$) is
$$j^-_\mu(x)=-{i\over 2\pi}\epsilon_{\mu\nu}\partial_\nu(\phi+{1\over 6M^2}
\partial_\alpha^2\phi +\cdots)\eqno(6.50)$$
With $\mu,\nu,\alpha= u,v$. In the notation used in the proof of the theorem
stated above, this formula corresponds to identify $J_{uv}=
-(i/2\pi)(\phi+\partial_v^2\phi/(6M^2) +\cdots)$, since $\phi$ does not
depend on $u$.
As expected the fermionic current at
lowest order is a closed 1-form whereas the higher orders
are total derivatives of $v$. $j^-_\mu(x)$ checks the general Eq. (6.43),
the higher orders vanishing only in the adiabatic limit. Note that
in the non adiabatic case the fermion number depends on the $v$-derivatives
of $\phi$ at $v=0$ and $v=T$, and hence does not agree with the zero
dimensional pseudoparity odd effective action.

\medskip
{\bf 7. Odd dimensions}
\medskip

In this section we will consider
the odd dimensional case in greater detail.
The main issue here is the quantum realization of parity which is a symmetry
at the classical level [31].
For odd $D$ and up to a similarity transformation, parity
corresponds to the transformation $\D=\ds +\Y(x)\to \ds -\Y(-x)$.
It turns $\lambda_n$ into $-\lambda_n$ in the eigenvalue equation (2.4).
As a consequence we can expect that divergent terms in the
action can break this symmetry.

In an inverse mass expansion of the action
we must look for invariance under $m\to -m$ and $\Dzero\to -\Dzero$, however
the formula (4.14) is not appropriate because it was derived for $m>0$.
The cut $\Gamma ={\rm R}^-$ in the integral $I_1$ breaks parity
explicitly, and indeed for negative $m$ the free spectrum $m\pm i|p|$ would
intersect $\Gamma$. In fact, there is no ray of minimal growth for both
positive and negative masses as required from Seeley's representation,
Eq. (2.7). To deal with this problem one can use some infrared
regulator such as including a factor $\theta(|p|-m_0)$ in the $p$ integral
and letting $m_0\to 0^+$.
However, a simpler procedure is to use dimensional regularization instead of
$\zeta$-function, which also maintains gauge invariance.
As always, both regularizations will coincide up to counterterms.
The procedure in this case is to compute the current
$\J=-\bra x|\D^{-1}|x\ket$ in a formal inverse mass expansion for a generic
dimension and afterwards to reconstruct the action. Of course we do this
using the Wigner transformation. One easily obtains the result
$$\J(x) = {1\over (4\pi)^{D/2}} \epsilon(m)^D \sum_{N=0}^\infty
a_N^J\,m^{D-N-1}\Oc_N(x;\Dzero) \eqno(7.1)$$
where $\D=\Dzero +m$, $\Oc(x)$ are the same coefficients as in Section 4,
and
$$\eqalign{
\epsilon(m) &= {m\over |m|}, \qquad m\in{\rm R} \cr
a_N^J &= (-1)^{N+1} \Gamma(\bar{N} - {1\over 2}D),\qquad
\bar{N}=\lfloor{N\over 2} +1\rfloor \cr} \eqno(7.2)$$
The terms of $\J$ in (7.1) which are ultraviolet finite
coincide with (4.18) (including the factor $\epsilon(m)^D$ which comes
from integral $I_1$ by making the momentum integral first). On the other
hand, the terms $N<D$ of the current depend on the regularization.
The coefficients $a_{N<D}^J$ have poles at even integer values of $D$.
Such poles give rise
to the logarithmic terms in (4.18). For odd $D$, the terms $N<D$ in
(7.1) give a finite
contribution proportional to $\epsilon(m)$, whereas in $\zeta$-function
they vanish for positive $m$. A more careful treatment would show that
in fact there is a step function factor $\theta(-m)$ in (4.18), so that both
regularizations differ only by a polynomial of $m$, since $m^k\theta(-m)=
m^k{1\over 2}-{1\over 2}m^k\epsilon(m)$. The sign function $\epsilon(m)$
does not necessarily vanish at $m=0$; its precise value is ambiguous and
further information is required.

The current in (7.1) is manifestly gauge covariant and also
parity covariant (for odd  $D$) thanks to the factor $\epsilon(m)$,
since both symmetries are respected by dimensional regularization and
the current turns out to be finite. In contrast, to compute the action directly
in dimensional regularization one would need a Seeley-like representation
$\Tr\log(\D) = -(2\pi i)^{-1}\Tr\int_\Gamma\dd z\log(z)(\D-z)^{-1}$, thus
manifestly breaking parity invariance. This situation resembles the method
of Leutwyler [12] and Ball [2], based on the integrability
obstruction of the covariant regularization of the formal variation
of the action.
We must then consider whether $\J$ is a consistent current and independent
of the choice of the splitting of $\D$ into $\Dzero$ and $m$, that is,
whether $\delta_m\J=0$. To answer these questions let us consider separately
the ultraviolet finite and divergent parts, $\J=\J_f+\J_d$, respectively.
The finite part coincides with the $\zeta$-function current and hence
it is consistent, indeed it is the variation of the action (4.14)
(after including the $\epsilon(m)$ factor). Also it is invariant under
$\delta_m$
by the same arguments as in Section 4, with the prescription of defining
$\epsilon(m)m^{D-N+1}$ as a derivative of the distribution
$\epsilon(m)\log|m|$ in the $m$ variable.

Consider now the divergent part of the current, $\J_d$ corresponding
to $N<D$, and odd $D$. From (4.12) it is immediate to check that
$\delta_m\J_d=0$ if $m\not=0$, that is, except for Dirac delta terms
coming from $\delta_m\epsilon(m)$. Hence
we can resum all these terms by pulling out the factor $\epsilon(m)$ and
setting $\Dzero\to \D$ and $m\to 0$ in the remainder, thus
$$\J_d = -{\epsilon(m)\over 2^D\pi^{(D-1)/2}}\Oc_{D-1}(x;\D)\,. \eqno(7.3)$$
Because $\bra\Oc_D\ket$ vanishes for odd $D$, it is not obvious whether
$\J_d$ is consistent. We can check that up to the $\epsilon(m)$ factor,
to be discussed below, the current appears to be consistent
for $D=1$ and $3$. Indeed, this can be accomplished by
writing the possible terms in the action and adjusting their coefficients.
In the case $D>1$, one can see that this cannot be done with just terms
of the form $\bra P(\D)\ket$, $P$ being a polynomial, and one must
resort to polynomials on $\D=\ds+\Y$ and $\Db=\ds-\Y$. For given $\J$
the solution is unique up to an absolute constant. We find
$$\eqalign{
(D=1)\qquad\qquad W_d &= -{\epsilon(m)\over 2}\int\dd x\,\tr\,(\Y) \cr
(D=3)\qquad\qquad W_d &= -{\epsilon(m)\over 8\pi}\int\dd^3 x\,\tr\,(
{1\over 12}\Y^3 +{1\over 4}\Y\gamma_\mu\Y\gamma_\mu\Y
+{1\over 2}\Y\ds\Y) \cr} \eqno(7.4)$$
In the $D=3$ case $\gamma_0\gamma_1\gamma_2=\sigma =\pm 1$ distinguishes
the two inequivalent irreducible representation of $\gamma_\mu$,
and $\Y=\thru{A} +M$,\  $A_\mu$, $M$ being Hermitian flavor matrices. Working
out the Dirac space algebra, the divergent parts of the current and the
action can be written more explicitly as
$$\eqalign{
\J_d &= {\epsilon(m)\over 8\pi}(2M^2+{1\over 2}\sigma\epsilon_{\mu\nu\alpha}
iF_{\mu\nu}\gamma_\alpha) \cr
W_d &= {\epsilon(m)\over 8\pi}\int\dd^3 x\,\tr\,
({4\over 3}M^3-i\sigma\epsilon_{\mu\nu\alpha}
(A_\mu\partial_\nu A_\alpha - {2\over 3}iA_\mu A_\nu A_\alpha)) \cr}
\eqno(7.5)$$
Here $\tr$ no longer acts on Dirac space. Up to the factor
$\epsilon(m)$, the pseudoparity odd term of the action, $W^-_d$,
is the correctly normalized Chern-Simons action
[31] and it is gauge invariant.

Finally let us consider the issue of whether $\J_d$ is invariant under
$\delta_m$, i.e. whether it is independent on the choice of the expansion
point $m$. The invariance is broken by the factor $\epsilon(m)$ in
(7.3), and $\delta_m\J_d= -(4\pi)^{-(D-1)/2}\Oc_{D-1}\delta(m)\delta m$
does not vanish as a distribution.
Still one can consider adding counterterms to cancel
the variation. The rationale is as follows. The ultraviolet divergent
integrals become well defined by applying $\partial/\partial m$ a sufficient
number of times, hence the ambiguity in the integral must be a polynomial
in $m$. By the same token, expressions which diverge as $m\to 0$ become
well defined after multiplying by $m$ a sufficient number of times, hence
the infrared ambiguities, i.e. the admissible counterterms,
consist of the distribution $\delta(m)$ and
its derivatives. Unlike the ultraviolet case, however, there are infrared
divergences to all orders in the inverse mass expansion.

For instance in the case $D=1$, if we change $\J_d = -{1\over 2}\epsilon(m)$
to $\J_d = -{1\over 2}\bra x|\epsilon(\D)|0\ket$, and expand in powers of
$\Dzero$, the difference are just infrared counterterms and now
$\delta_m\J_d=0$ explicitly. Unfortunately this current
is no longer gauge invariant
nor consistent. So we must consider simultaneously the three conditions
of $\delta_m$ invariance, gauge invariance and consistency of the current
after introducing infrared counterterms. Considering again the $D=1$ case
one can see that these conditions are not compatible. Indeed let
$\gamma_0=i\sigma$, $\Y=M+\gamma_0 A_0$, $M,A_0\in$R.
$\delta_m$ invariance implies that $\J_d$ must depend only on the complex
number $\Y$, consistency
requires that $\J_d$ must be an analytic function of $\Y$ and finally
gauge invariance requires $\J$ to be independent of $\Im\Y$, therefore
$\J_d$ must be a constant, which contradicts $\J_d = -{1\over 2}\epsilon(m)$
for $\Y=m$. Also, this solution would break parity, which requires
the current to be an odd function of $\Y$.

In conclusion, if we insist that the current and the action must be
gauge invariant we must choose between two possibilities. First, take
$m$ and $\Dzero$ as independent variables. In this case the actions
(7.4) are simultaneously gauge and parity invariant. And second, that
the action depends only on $\D$. In this case we obtain two different
actions by taking $m$ positive or negative both of which are gauge
invariant but have a parity anomaly. This is what happens also in the
massless case [31]. The choice between the two possibilities depends
on the physical problem at hand, for instance whether the problem admits
a natural definition of $m$ or not.
Note that the ultraviolet finite part of the action
was independent of $m$.

\medskip
{\bf Acknowledgments}
\smallskip
{\sl
This work has been partially supported by the DGICYT under contract
PB92-0927 and the Junta de Andaluc\'{\i}a (Spain), as well as
FOM and NWO (The Netherlands). One of us (E.R.A.) acknowledges NIKHEF-K
for hospitality.
}
\vfill\eject

\medskip
{\bf Appendix A. Seeley-DeWitt coefficients for Dirac operators to all
orders}
\medskip
In the case of even order, the Seeley-DeWitt coefficients for $\D$ admit
a simple form.
The general pattern can be illustrated by inspection of the first coefficients
up to $\Oc_6(\D)$

$$\eqalign{
\Oc_0(\D) &= 1 \cr
\Oc_2(\D) &=  \D^2+\A^2 \cr
\Oc_4(\D) &=  {1\over 2}\D^4 +{1\over 3}(\D^2\A^2+\A\D^2\A + \A^2\D^2)
             + {1\over 6} \A^4 \cr
\Oc_6(\D) &=  {1\over 6}\D^6 +
               {1\over 12}(\D^4\A^2+\D^2\A\D^2\A +
\D^2\A^2\D^2 +\A\D^4\A+\A\D^2\A\D^2 \D^4\A^2\D^4) \cr
&+{1\over 30}(\D^2\A^4 + \A\D^2\A^3+ \A^2\D^2\A^2+ \A^3\D^2\A + \A^4\D^2)
 + {1\over 90} \A^6 \cr
}\eqno({\rm A}.1)$$
Here, each monomial of degree $2m$ in $\A$, stands for the $(2m-1)!!$
terms obtained by all possible contractions in the Lorentz indices involving
the $2m$ $\A_\mu$ vectors. (Compare with Eq. (4.7) for $\Oc_2$ and $\Oc_4$).
Note that $\D^2$ does not commute with the symbol $\A$ and also that
the cyclic property cannot be applied.

One realizes that these coefficients depend only on $\A_\mu$ and $\D^2$.
This is in fact a general rule which can be inferred from a direct
comparison of inverse mass and Heat Kernel expansions (see Eq. (B.7)).
Moreover, we note that the $\left(\matrix{n+2m \cr n\cr }\right)$
monomials of degree $n$ in $\D^2$ and $2m$ in $\A$, with $2n+2m=N$,
appear in all combinations and with the same coefficient $a_{N,n}$,
namely
$$a_{N,n} = {2^{N/2 -n}\over(N-n)!} \eqno({\rm A}.2)$$
In summary, the final formula for $\Oc_N(\D)$ reads
$$\Oc_N(\D) = \sum_{n=0}^{N/2}a_{N,n}[\D^{2n}\A^{N-2n}] \eqno({\rm A}.3)$$
Where $[\D^{2n}\A^{N-2n}]$ stands for the sum of the
$\left(\matrix{N-n \cr n\cr }\right)$ terms obtained by all possible
monomials of degree $n$ in $\D^2$ and $N-2n$ in $\A$.
As expected, the coefficients do not depend explicitly on the space time
dimension $D$.
In fact, if one assumes the validity of Eq. ({\rm A}.3), the coefficients can
be completely determined by taking particular cases.
For instance, the relative weight
of the coefficients is determined by gauge invariance, and the global
normalization can be fixed by either going to the $D=0$ case or
by taking $\D$ to be a c-number.
Finally, we mention that the odd order coefficients can be derived
by direct use of the formula
$\delta_m\Oc_N = -\epsilon_{N-D}\Oc_{N-1}\delta m $.

The previous result can straightforwardly be applied to compute the chiral
anomaly in $D$ dimensions for a purely vector external field. Defining
the field strength tensor $iF_{\mu\nu}=[\Dc_\mu,\Dc_\nu]$, one finds
$${\cal A}_A = -{1\over (D/2)!}
\epsilon_{\mu_1\cdots\mu_D}\left\langle 2i\alpha
F_{\mu_1\mu_2}\cdots F_{\mu_{D-1}\mu_D}\right\rangle
\eqno({\rm A}.4)$$
where $\langle\ \rangle$ was defined in Eq. (4.15) and $\tr$ includes
flavor space only.

\vskip1cm
{\bf Appendix B. Inverse mass and Heat Kernel expansions for
Dirac operators in even dimensions}
\medskip

In even dimensions a straightforward relation between the well-known
Heat Kernel expansion and the inverse mass expansion of Section 4
can be established. Our result is a reminiscent of the formal determinantal
relation $\Det(\D)\Det(-\D) = \Det(-\D^2)$. Actually, what
is found is that this relation holds for the $\zeta$-function
regularization of the determinants at each order in an inverse mass
expansion, where the l.h.s. stands for the Wigner transformation result and
the r.h.s. for the Heat Kernel expansion.

Let $\Delta$ be a
positive definite second order differential operator of the form
$$\Delta = -\hat{\cal D}_\mu^2 + Y(x),\quad\quad
i\hat{\cal D}_\mu = i\partial_\mu+B_\mu(x)  \eqno({\rm B}.1)$$
The heat kernel $\bra x|e^{-\tau\Delta}|x\ket$ is
ultraviolet finite for $\tau>0$ and it admits an asymptotic
expansion [29] around $\tau=0$
$$ \bra x|e^{-\tau\Delta}|x\ket
= (4\pi\tau)^{-D/2}\sum_{n=0}^\infty\tau^na_n(x) \eqno({\rm B}.2)$$
where the Seeley-DeWitt coefficients
$a_n(x)$ are covariant local polynomials in $i\hat{\cal D}_\mu$
and $Y$ of degree $p$, $q$ respectively with $2p+q=n$. This
coefficients can be written in a dimension independent way [2].
If we proceed formally, relying on the good behavior of the required
analytical continuations, we find
$$\eqalign{
\bra x|\D^s|x\ket &= {(-1)^{-s/2}\over \Gamma(-{s\over 2})}
 \int_0^\infty\dd\tau \tau^{-s/2-1} e^{\tau m^2}
\bra x|e^{-\tau\Delta}|x\ket \cr
 \Delta &= -\D^2+m^2  = -\Dzero^2-2m\Dzero \cr
}\eqno({\rm B}.3)$$
Now, by using the heat kernel
expansion, the $\tau$ integral can be carried out order by order
(for $m^2<0$ and then analytically continued to $m^2>0$). In
this way we obtain another representation of the $\zeta$-function
$$\bra x|\D^s|x\ket = \sum_{n=0}^\infty(-1)^{{D\over 2}-n}
{m^{D+s-2n}\over(4\pi)^{D/2}} {\Gamma(n-{D+s\over
2})\over\Gamma(-{s\over 2})}a_n(x) \eqno({\rm B}.4)$$
This is to be compared to the expansion obtained from the
Wigner transformation approach
$$\bra x|\D^s|x\ket = \sum_{N=0}^\infty
{m^{D+s-N}\over(4\pi)^{D/2}} {\Gamma(s+1)\over\Gamma(s+1-N+D)}
\zeta_{N-D}\Oc_N(x) \eqno({\rm B}.5)$$
The comparison however is not immediate because we must first of
all reexpand $a_n(x)$ in powers of $m$, or equivalently in
powers of $\Dc_\mu$ and $\X$, where $\Dzero=\Dcs+\X$. For a given
order $N$, $\Oc_N(x)$ generically gets contributions from all the terms
$a_n(x)$ with $n\ge {1\over 2}N$. This is because $a_n(x)$ is
made out of $i\hat{\cal D}_\mu$ and $Y(x)$ which are of zeroth order in
$\Dc_\mu$ and $\X$, namely
$$\eqalign{
i\hat{\cal D}_\mu &= \Dc_\mu-{1\over 2}\{\gamma_\mu,\X+m\} \cr
Y &= m^2-(\X+m)^2-{1\over 2}[\gamma_\mu,[\Dc_\mu,\X]] - {1\over 4}
\{\gamma_\mu,\X+m\}^2-{1\over 2}\sigma_{\mu\nu}[\Dc_\mu,\Dc_\nu] \cr
}\eqno({\rm B}.6)$$
Nevertheless, whenever $\Oc_N(x)$ requires only a finite number of heat kernel
terms we can check that both expressions coincide. In particular
choosing $D$ even, $s=m=0$, we find the following identity
$(4\pi)^{D/2}\bra x|\D^0| x\ket =\Oc_D(x;\D) = a_{D/2}(x;\Delta=-\D^2)$, which
holds regardless of the way the coefficients are written.
If both coefficients are expressed in a dimension independent form,
one has the further identity, for even dimensions
$$\Oc_{2n}(x;\D)= a_n(x;\Delta = -\D^2), \quad\quad n =
0,1,2,\dots \eqno({\rm B}.7)$$
This can be better checked for lowers orders rewriting Eq.
({\rm B}.6) (for $m=0$) as
$$i\hat{\cal D}_\mu= -\A_\mu,\quad\quad
Y=-\D^2-\A_\mu^2 \eqno({\rm B}.8)$$
Using well known expressions for $a_0$, $a_1$, $a_2$ [2], we
reproduce $\Oc_0$, $\Oc_2$ and $\Oc_4$. An interesting consequence
of $\Oc_D(x;\D) = a_{D/2}(x;\Delta=-\D^2)$ is that, although an anomaly
calculation using $\D^2$ as regulator might be questionable in principle
for non normal $\D$ or non positive $-\D^2$,
such a procedure turns out to be justified a posteriori.

\vfill\eject
{\bf Appendix C. Explicit form of the axial anomaly in four dimensions}
\medskip

\def\Dm{i{\cal D}_\mu}
\def\Dn{i{\cal D}_\nu}
\def\Ds{i{\cal D}_\sigma}
\def\Dr{i{\cal D}_\rho}
\def\Am{A_\mu}
\def\An{A_\nu}
\def\As{A_\sigma}
\def\Ar{A_\rho}
\def\Tmn{T_{\mu\nu}}
\def\Tab{T_{\alpha\beta}}
\def\Tma{T_{\mu\alpha}}
\def\Tna{T_{\nu\alpha}}
\def\Tmb{T_{\mu\beta}}
The most general spinor structure for the Dirac operator in four
Euclidean dimensions is given by
$$\D = i\!\thru{\cal D}+S+i\gamma_5
P+\thru A\!\gamma_5 + {1\over 2}i\sigma_{\mu\nu}\Tmn-{1\over
2}\sigma_{\mu\nu}\gamma_5 \Tmn^\prime \eqno({\rm C}.1)$$
with $ i{\cal D}_\mu = i\partial_\mu + V_\mu $,
$\sigma_{\mu\nu} = {1\over 2} [\gamma_\mu ,\gamma_\nu] $
and  $ T^*_{\mu\nu} = {1\over 2}
\epsilon_{\mu\nu\rho\sigma} T_{\rho\sigma} $.
If the Hamiltonian $\gamma^0\D$ is Hermitian in Minkowski space, and
assuming that the fields transform as usual under the Wick rotation,
the terms $S$, $\thru A\!\gamma_5$ and ${1\over 2}i\sigma_{\mu\nu}\Tmn$
are Hermitian whereas $i\!\!\thru{\cal D}$, $i\gamma_5 P$ and $
{1\over 2}\sigma_{\mu\nu}\gamma_5\Tmn^\prime$ are antihermitian.
This allows to distinguish between the two tensor terms. However, as our
expressions will not depend on the hermiticity properties of $\D$ we
can reabsorb $\Tmn^\prime$ into $\Tmn$, i.e., $\Tmn-i\Tmn^{\prime *}
\to \Tmn$.

In the following expressions, $\langle\ \ \ \rangle$ stands for $\int{\rm
d}^4 x\,{\rm tr}\,{1\over 4\pi^2} \alpha(x)$ and tr refers to flavor
only. The axial anomaly is obtained adding up all the terms.
 $$\eqalign{
{\cal A}[V^4] &= {1\over 4}i\epsilon_{\mu\nu\rho\sigma}\langle\
[\Dm,\Dn][\Dr,\Ds] \ \ \rangle \cr
{\cal A}[A^4] &= -{1\over 3} i\epsilon_{\mu\nu\rho\sigma} \langle\
\Am \An \Ar \As \ \rangle \cr
{\cal A}[V^2A^2] &= -i\epsilon_{\mu\nu\rho\sigma}\langle\
{1\over 6}\{[\Dm,\Dn],\Ar\As\} + {2\over 3}
\Am[\Dn,\Dr]\As -{1\over 3}[\Dm,\An][\Dr,\As]\
\rangle \cr
{\cal A}[V^3A] &= i\langle\  [\Dm,[\Dm,[\Dn,\An]]] -
[\Dm,[\Dn,[\Dm,\An]]] +{1\over
3}[\Dn,[\Dm,[\Dm,\An]]] \ \rangle \cr
{\cal A}[VA^3] &= i\langle\  [\Dm,\An]\An \Am + \Am
\An[\Dm,\An] + {1\over 3}\An \An[\Dm,\Am] \cr
&+ {1\over 3}[\Dm,\Am]\An \An -{1\over 3}[\Dm,\An]\Am
\An -{1\over 3}\An \Am[\Dm,\An] \cr
&+ {2\over 3}\An[\Dm,\Am]\An +{2\over
3}\Am[\Dm,\An]\An +{2\over 3}\An[\Dm,\An]\Am \ \rangle
\cr
{\cal A}[S^3P] &= \langle\  -S\{S,P\}S + {5\over 3}\{S^3,P\} \ \rangle \cr
{\cal A}[SP^3] &= \langle\   P\{S,P\}P + \{S,P^3\} \ \rangle \cr
{\cal A}[SPV^2] &= \langle\ \{[\Dm,P],[\Dm,S]\} + {1\over
3}\{[\Dm,[\Dm,S],P\} +\{[\Dm,[\Dm,P]],S\} \ \rangle \cr
{\cal A}[S^2VA] &= i\langle\ \{S^2,[\Dm,\Am]\} +S\Am[\Dm,S] + [\Dm,S]\Am S \cr
&+ {5\over 3}[\Dm,S]S\Am +{5\over 3}\Am S[\Dm,S]
+{4\over 3}S[\Dm,S]\Am +{4\over 3}\Am[\Dm,S]S \ \rangle \cr
{\cal A}[P^2VA] &= i\langle\ [[\Am,P],[\Dm,P]] +\{[\Dm,\Am],P^2\}
\ \rangle \cr
{\cal A}[SPA^2] &= \langle\ [\Am \Am,[S,P]] + \{\{\Am,S\},\{\Am,P\}\}
\ \rangle \cr
}$$
$$\eqalign{
{\cal A}[T^4] &= i\langle\  {1\over
12}\Tmn \Tab \Tab \Tmn^* -{1\over 4}\Tmn\Tab\Tab^*\Tmn
-{1\over 3}\Tmn [\Tab,\Tmb]\Tna^* \ \rangle \cr
{\cal A}[ST^3] &= {1\over 3}\langle\ \Tmn^*\Tma\Tna S
+S\Tma\Tna\Tmn^* +
\Tma\{S,\Tmn^*\}\Tna \ \rangle \cr
{\cal A}[PT^3] &= i\langle\ \Tma\{P,\Tmn \}\Tna -{1\over
3}\Tmn \Tma\Tna P -{1\over
3}P\Tma\Tna\Tmn \ \rangle \cr
{\cal A}[S^2T^2] &= i\langle\ {5\over 6}\Tmn S^2\Tmn^* -{1\over
2}S\Tmn\Tmn^*S
+{1\over 6}S\{S,\Tmn^*\}\Tmn  +{1\over
6}\Tmn \{S,\Tmn^*\}S \ \rangle \cr
{\cal A}[P^2T^2] &= {1\over 2}i\langle\
\Tmn P^2\Tmn^*+P\Tmn \Tmn^*P
+P\{P,\Tmn \}\Tmn^* + \Tmn^*\{P,\Tmn \}P \ \rangle
\cr
{\cal A}[SPT^2] &= \langle\ {1\over 6}S\Tmn \Tmn P + {1\over 6}P\Tmn
\Tmn S -{1\over 2}\Tmn \{S,P\}\Tmn  +{1\over
6}P\{S,\Tmn \}\Tmn  \cr
&+ {1\over 6}\Tmn \{S,\Tmn \}P -{1\over 2}S\{P,\Tmn \}\Tmn
-{1\over 2}\Tmn \{P,\Tmn \}S \ \rangle \cr
{\cal A}[V^2T^2] &= i\langle\
\{[\Dm,[\Dn,\Tna ]],\Tma^*\} +{1\over
3}\{[\Dm,[\Dn,\Tna^*]],\Tma\} \cr
&+ {1\over 3} \{[\Dm,\Tma],[\Dn,\Tna^*]\}
+{1\over 3} [\Dm,\Tab][\Dm,\Tab^*]\
\rangle \cr
{\cal A}[A^2T^2] &= i\langle\ {1\over 2}\Tab \Am \Am
\Tab^* +{1\over 2}\Am \Tab \Tab^* \Am
+{1\over 3}\Tma[\Am,\An]\Tna^* \cr
&+{1\over 3} \Am(\Tma \Tna^* -
\Tna \Tma^*)\An -{1\over
3}\Tna \{\Am,\Tma^*\}\An \cr
&-{1\over 3}\An\{\Am,\Tma^*\}\Tna  - {1\over
3}\Tna^*[\Am,\Tma]\An +{1\over
3}\An[\Am,\Tma]\Tna^*\ \rangle \cr
{\cal A}[VAT^2] &= -i\langle\
{1\over 6}\{\Tab \Tab ,[\Dm,\Am]\}
-{1\over 3}\{\{\Tma,\Tna \},[\Dm,\An]\}
+{1\over 3}[\Dm,\Tma]\Tna \An \cr
&+{1\over 3}\An \Tna [\Dm,\Tma]
-{1\over  6}[\Dm,\Tab ]\Tab \Am
-{1\over 6}\Am \Tab [\Dm,\Tab ] \cr
&-{1\over 3}[\Dm,\Tna ]\Tma \An
-{1\over 3}\An \Tma[\Dm,\Tna ]
-{4\over 3}\Tma[\Dn,\Tna]\Am \cr
&-{4\over 3}\Am[\Dn,\Tna]\Tma
-{1\over 3}[\Dm,\Tma]\An \Tna
-{1\over 3}\Tna\An[\Dm,\Tma] \cr
&+{1\over 3}[\Dm,\Tna]\An\Tma
+{1\over 3}\Tma\An[\Dm,\Tna]
+{1\over 6}[\Dm,\Tab]\Am\Tab \cr
&+{1\over 6}\Tab\Am[\Dm,\Tab]
+{2\over 3}\Tab[\Dm,\Am]\Tab \ \rangle \cr
{\cal A}[SV^2T] &= -\langle\ {2\over 3}S[\Dm,\Dn]\Tmn^*
+ {2\over 3}\Tmn^*[\Dm,\Dn]S \cr
&+{1\over 6}\{\{S,\Tmn^*\},[\Dm,\Dn]\}
+{1\over 3}[[\Dm,\Tmn^*],[\Dn,S]] \ \rangle \cr
{\cal A}[SA^2T] &= -{1\over 3}\langle\ \Tmn^*\Am\An S+S\Am\An\Tmn^*
+\Am\{S,\Tmn^*\}\An
-3\Tmn^*\{S,\An\}\Am \cr
&+ 3\Am\{S,\An\}\Tmn^*
-S\{\Am,\Tmn^*\}\An +\An\{\Am,\Tmn^*\}S \ \rangle \cr
}$$
$$\eqalign{
{\cal A}[PV^2T] &= i\langle\ {1\over 2}\{\{\Tmn ,P\},[\Dm,\Dn]\}
-[[\Dm,\Tmn ],[\Dn,P]] \ \rangle \cr
{\cal A}[PA^2T] &= i\langle\ {1\over 3}\Tmn\Am\An P + {1\over 3}P\Am\An\Tmn
-\Am\{P,\Tmn \}\An +\Tmn [\Am,P]\An \cr
&- \Am[\An,P]\Tmn  -{1\over 3}P[\Am,\Tmn ]\An +{1\over
3}\Am[\An,\Tmn ]P \ \rangle \cr
{\cal A}[SVAT] &= \langle -[\Dm,S]\An \Tmn +\Tmn \An[\Dm,S] -[\Dm,\An]S\Tmn
+\Tmn
S[\Dm,\An] \cr
&+{1\over 3}[\Dm,S]\Tmn \An -{1\over 3}\An \Tmn[\Dm,S] -{1\over
3}[\Dm,\An]\Tmn S \cr
&+{1\over 3}S\Tmn[\Dm,\An] +{1\over 3}[\Dm,\Tmn]S\An -{1\over
3}\An S[\Dm,\Tmn] \cr
&+{1\over 3}[\Dm,\Tmn]\An S -{1\over 3}S\An[\Dm,\Tmn]
+{2\over 3}\Tmn[\Dm,S]\An \cr
&- {2\over 3}\An[\Dm,S]\Tmn +{2\over 3}\An[\Dm,\Tmn]S
-{2\over 3}S[\Dm,\Tmn]\An \cr
&+{2\over 3}\Tmn[\Dm,\An]S -{2\over 3}S[\Dm,\An]\Tmn \
\rangle \cr
{\cal A}[PVAT] &= i\langle\ [\Dm,P]\An \Tmn^* -\Tmn^*\An[\Dm,P]
+[\Dm,P]\Tmn^*\An \cr
&-\An \Tmn^*[\Dm,P] -{1\over 3}[\Dm,\An]\Tmn^*P +{1\over 3}P\Tmn^*[\Dm,\An] \cr
&-[\Dm,\An]P\Tmn^* +\Tmn^*P[\Dm,\An] +[\Dm,\Tmn^*]P\An \cr
&- \An P[\Dm,\Tmn^*]-{1\over 3}[\Dm,\Tmn^*]\An P +{1\over
3}P\An[\Dm,\Tmn^*] \cr
&+ 2\Tmn^*[\Dm,P]\An -2\An[\Dm,P]\Tmn^* +{2\over 3}P[\Dm,\An]\Tmn^* \cr
&- {2\over 3}\Tmn^*[\Dm,\An]P +{2\over 3}P[\Dm,\Tmn^*]\An -{2\over
3}\An[\Dm,\Tmn^*]P \ \rangle \cr
} $$
\vskip1cm
{\bf Appendix D. Explicit form of the counterterms in four dimensions}
\vskip 12pt

For all the terms which do not contain $\Dm$ nor $\Am$, one can find a
compact expression for their counterterms, namely
$$W[S,P,T] = -{1\over 24}
\big\langle
\M^4+\M^2\gamma_\mu\M^2\gamma_\mu +2\M^3\gamma_\mu\M\gamma_\mu
\big\rangle
\eqno({\rm D}.1)$$
where here $\M(x) =S+i\gamma_5P+{1\over 2}i\sigma_{\mu\nu}\Tmn$
and $\delta_A\M(x)=-\{i\alpha(x)\gamma_5,\M(x)\}$.
The rest of the counterterms are given by

$$\eqalign{
W[A^4+\cdots] &= -2\big\langle
[\Dm,\Am][\Dn,\An]-[\Dm,\An][\Dn,\Am] \cr
&+{1\over 3}[\Dm,\An][\Dm,\An]-{4\over 3}\Am\Am\An\An+{2\over
3}\Am\An\Am\An\big\rangle \cr
W[S^2V^2+\cdots] &= 4\big\langle
{1\over 3}[\Dm,S][\Dm,S]
+{1\over 2}[\Dm,P][\Dm,P]  \cr
&+S^2\Am\Am-P\Am P\Am-i[\Dm,S]\{P,\Am\}\big\rangle \cr
W[V^2T^2+\cdots] &= 4\big\langle-{1\over
6}[\Am,\Tma][\An,\Tna]+{1\over
6}\Am\Am\Tab\Tab \cr
&-{1\over 3}[\Dm,\Tma][\Dn,\Tna]
+{1\over 6}[\Dm,\Tna][\Dn,\Tma] \cr
&-{1\over 6}[\Dm,\Am]\Tab\Tab^*
+{1\over 3}[\Dm,\Tma]\{\An,\Tna^*\}\big\rangle \cr
W[SV^2T+\cdots] &= -4i\big\langle
-{1\over 3}\{S,\Tmn\}\Am\An
+{4\over 3}S\Am\Tmn\An +{1\over 2}i[\Dm,\Dn]\{P,\Tmn\} \cr
&+{1\over 3}[\Dm,S][\An,\Tmn^*]
-{2\over 3}[\Dm,\An][S,\Tmn^*] -i[\Dm,\Tmn][P,\An]
\big\rangle \cr}$$

\vfill\eject
\centerline{\bf REFERENCES}
\vskip0.5cm
\item{[1]} P. Ramond, Field theory: a modern primer, (Addison Wesley,
	1990).
\item{[2]} R. Ball, Phys. Rep. 182 (1989) 1.
\item{[3]} A.P. Balachandran, G. Marmo and C.G. Trahern, Phys. Rev.
 D25 (1982) 2713.
\item{[4]} S.K. Hu, B.L. Young and D.W. McKay, Phys. Rev. D30 (1984) 836.
\item{[5]} A. Andrianov and L. Bonora, Nucl. Phys. B223 (1984) 232.
\item{[6]} J.L. Petersen, Acta. Phys. Pol. B16 (1985) 271.
\item{[7]} J. Gasser and H. Leutwyler, Ann. Phys. 158 (1984) 142.
\item{[8]} M. Reuter, Phys. Rev. D31 (1985) 1374.
\item{[9]} L.H. Chan, Phys. Rev. Lett. 57 (1986) 1199.
\item{[10]} D.G.C. McKeon and T.N. Sherry, Ann. Phys. 218 (1992) 325.
\item{[11]} K. Fujikawa, Phys. Rev. D29 (1984) 285.
\item{[12]} H. Leutwyler, Phys. Lett. B152 (1985) 78.
\item{[13]} S.L. Adler, Phys. Rev. 177 (1969) 2426.
\item{[14]} J. Bell and R. Jackiw, Nuovo Cimento A60 (1969) 47.
\item{[15]} W.A. Bardeen, Phys. Rev. 184 (1969) 1848.
\item{[16]} S.L. Adler and W.A. Bardeen, Phys. Rev. 182 (1969) 1517.
\item{[17]} J. Wess and B. Zumino, Phys. Lett. B37 (1971) 95.
\item{[18]} K. Fujikawa, Phys. Rev. D21 (1980) 2848.
\item{[19]} J. Goldstone and F. Wilczek, Phys. Rev. Lett. 47 (1981) 986.
\item{[20]} E. Witten, Nucl. Phys. B223 (1983) 422.
\item{[21]} R.E. Gamboa Sarav\'{\i}, M.A. Muschietti, F.A. Schaposnik
	and J.E. Solomin, Ann. Phys. (N.Y.) 157 (1984) 360.
\item{[22]} S.W. Hawking, Comm. Math. Phys. D55 (1977) 133.
\item{[23]} R.T. Seeley, Amer. Math. Soc. Proc. Symp. Pure Math.
	10 (1967) 288.
\item{[24]} E.P. Wigner, Phys. Rev. 40 (1932) 749.
\item{[25]} M. Hillery, R.F. O$^\prime$Connel, M.O. Scully
	and E.P. Wigner, Phys. Reports, 106 (1984) 121.
\item{[26]} W.A. Bardeen and B. Zumino, Nucl. Phys. B244 (1984) 421.
\item{[27]} R.E. Gamboa Sarav\'{\i}, M.A. Muschietti, and J.E. Solomin,
	Comm. Math. Phys. 89 (1983) 363.
\item{[28]} L. H\"ormander, The analysis of linear partial
	differential operators, vol. III, (Springer-Verlag, Berlin, 1985).
\item{[29]} P. Gilkey, Index theorems and the Heat equation (Publish or
	Perish, Berkeley, 1975).
\item{[30]} E.P. Wigner, Phys. Rev. 46 (1934) 1002;
	J.G. Kirkwood, Phys. Rev. 44 (1933) 31.
\item{[31]} L. \'Alvarez-Gaum\'e, S. Della Pietra and G. Moore,
	Ann. Phys. (N.Y.), 163 (1985) 288.
\item{[32]} R.E. Gamboa Sarav\'{\i}, M.A. Muschietti, F.A. Schaposnik and
	J.E. Solomin, Jour. Math. Phys. 26 (1985) 2045.
\item{[33]} T.E. Clark and S.T. Love, Nucl. Phys. B223 (1983) 135.
\item{[34]} C. Lee and J. Minn, Phys. Rev. D35 (1987) 1872.
\item{[35]} J. Bijnens and J. Prades, Phys. Lett. B320 (1993) 130.
\item{[36]} A. Dhar, R. Shankar and S.R. Wadia, Phys. Rev. D31 (1985) 3256.
\item{[37]} D. Ebert and H. Reinhardt, Nucl. Phys. B271 (1986) 188.
\item{[38]} E. Ruiz Arriola and L.L. Salcedo, Mod. Phys. Lett. A8 (1993)
2061.
\item{[39]} J. Caro, E. Ruiz Arriola and L.L. Salcedo, Granada Preprint,
UG-DFM-32/94.
\item{[40]} M. Abramowitz and I.A. Stegun, Handbook of
mathematical functions, (Dover, New York, 1972).
\item{[41]} Y. Brihaye, N.K. Pak and P. Rossi, Phys. Lett. B149 (1984) 191.
\item{[42]} \"O. Kaymak{\c c}alan, S. Rajeev and J. Schechter, Phys. Rev.D30
(1984) 594.
\item{[43]} N.K. Pak and P. Rossi, Nucl. Phys. B250 (1985) 279.
\item{[44]} J. Ma\~nes, Nucl. Phys. B250 (1985) 369.
\item{[45]} E. D'Hoker and E. Farhi, Nucl. Phys. B248 (1984) 59.
\item{[46]} R. Ball and H. Osborn, Nucl. Phys. B263 (1986) 245.

\vfill\eject

{\bf Table captions}\medskip

\item{Table 4.1} Several useful particular values for the coefficients
involved in the inverse mass expansion (4.2) and defined in Eqs. (4.5) and
(4.6). The coefficients for the inverse mass expanded effective
action Eq. (4.14) and the effective current Eq. (4.18) and (4.19) are also
given.
In the odd-dimensional case, the coefficients $\zeta_K$, $\alpha_K^W$,
$\beta_K^W$, $\alpha_K^J$ and $\beta_K^J$ have to be multiplied with an extra
$\sqrt\pi$ factor.
\medskip
\item{Table 4.2} Explicit formulas for the effective action inverse mass
coefficients, $\alpha_K^W$ and $\beta_K^W$ (see formula (4.14)) in terms
of $\zeta_K$ as given by Eqs. (4.5) and (4.6).

\vfill\eject

\hsize 18truecm


$$\hbox{\vbox{\tabskip=0pt \offinterlineskip
\def\strut{\hbox{\vrule height 18pt depth 8pt width 0pt}}
\halign{
\strut  \vrule# \tabskip= 1em plus2em
& \hfil$#$
& \vrule#
& \hfil$#$ & \hfil$#$ & \hfil$#$
& \hfil$#$ & \hfil$#$ & \hfil$#$
& \vrule#
& \hfil$#$ & \hfil$#$ & \hfil$#$
& \hfil$#$ & \hfil$#$ & \hfil$#$
& \vrule#\tabskip 0pt \cr
\noalign{\hrule}
& & \omit  & \multispan6 \hfil\hbox{even $D$}\hfil  & \omit &
\multispan6 \hfil\hbox{odd $D$}\hfil &\cr

& K
&\omit & \epsilon_K & \zeta_K & \alpha_K^W & \beta_K^W & \alpha_K^J & \beta_K^J
&\omit & \epsilon_K & \zeta_K & \alpha_K^W & \beta_K^W & \alpha_K^J & \beta_K^J
& \cr\noalign{\hrule}

& -5
&& -5 & 24 & {137\over 300} & -{1\over 5} & {25\over 12} & -1
&& 2 & 0 & 0 & 0 & 0 & 0 &
\cr
& -4
&& 2 & 12 & {25\over 24} &-{1\over 2} & {11\over 3} & -2
&& -4 & 0 & 0 & 0 & 0 & 0 &
\cr
& -3
&& -3 & -4 & -{11\over 9} & {2\over 3} & -3 & 2
&& 2 & 0 & 0 & 0 & 0 & 0 &
\cr
& -2
&& 2 & -2 & -{3\over 2} & 1 & -2 & 2
&& -2 & 0 & 0 & 0 & 0 & 0 &
\cr
& -1
&& -1 & 2 & 2 & -2 & 0 & -2
&& 2 & 0 & 0 & 0 & 0 & 0 &
\cr
& 0
&& 2 & 1 & 0 & -1 & -1 & 0
&& 0 & -1 & 0 & 1  & 1 & 0 &
\cr
& 1
&& 1 & 1 & -1 & 0 & 1 & 0
&& 2 & -{1\over 2} & {1\over 2} & 0 & -{1\over 2} & 0 &
\cr
& 2
&& 2 & {1\over 2} & {1\over 2} & 0 & -1 & 0
&& 2 & -{1\over 4} & -{1\over 4} & 0 & {1\over 2} & 0 &
\cr
& 3
&& 3 & {1\over 6} & -{1\over 3} & 0 & 1 & 0
&& 2 & -{1\over 8} & {1\over 4} & 0 & -{3\over 4} & 0 &
\cr
& 4
&& 2 & {1\over 12} & {1\over 2} & 0 & -2 & 0
&& 4 & -{1\over 32} & -{3\over 16} & 0 & {3\over 4} & 0 &
\cr
& 5
&& 5 & {1\over 60} & -{2\over 5} & 0 & 2 & 0
&& 2 & -{1\over 64} & {3\over 8} & 0 & -{15\over 8} & 0 &
\cr\noalign{\hrule}
}}}$$
\nobreak\centerline{\hbox{Table 4.1}}

\medskip

$$\hbox{\vbox{\offinterlineskip
\def\strut{\hbox{\vrule height 18pt depth 8pt width 0pt}}
\hrule
\halign{
\strut\vrule# \tabskip 0.1in
& \hfil$#$\hfil & \hfil$#$\hfil & \hfil$#$\hfil & \hfil$#$\hfil &
\vrule#\tabskip 0.0in
\cr
& & K<0 & K=0 & K>0 & \cr\noalign{\hrule}
& \alpha^W_K & {1\over\Gamma(-K+1)}\zeta_K\sum_{n=1}^{-K}n^{-1} & 0 &
(-1)^K\Gamma(K)\zeta_K & \cr\noalign{\hrule}
& \beta^W_K   & -{1\over\Gamma(-K+1)}\zeta_K & -\zeta_0 & 0 & \cr
\noalign{\hrule}}}}$$
\nobreak\centerline{\hbox{Table 4.2}}

\bye